\documentclass[journal]{IEEEtran}
\usepackage{amsmath}
\usepackage{amsfonts}
\usepackage{amsthm}
\usepackage{algorithmic}
\usepackage{algorithm}
\usepackage{array}
\usepackage{textcomp}
\usepackage{stfloats}
\usepackage{url}
\usepackage{verbatim}
\usepackage{graphicx}
\usepackage{cite}
\usepackage{xcolor}
\usepackage{setspace}
\usepackage{epstopdf}
\usepackage{bm}
\usepackage{amssymb}
\allowdisplaybreaks

\begin{document}
\title{Model Predictive Control Enabled UAV Trajectory Optimization and Secure Resource Allocation}
\author{Zhendong Li, Chang Su, Zhou Su, Haixia Peng, Yuntao Wang, Wen Chen and Qingqing Wu
\thanks{Z. Li, C. Su and H. Peng are with the School of Information and Communication Engineering, Xi'an Jiaotong University, Xi'an 710049, China (email: lizhendong@xjtu.edu.cn, suchang@stu.xjtu.edu.cn, haixia.peng@xjtu.edu.cn). }
\thanks{Z. Su and Y. Wang are with the School of Cyber Science and Engineering, Xi'an Jiaotong University, Xi'an 710049, China (email: zhousu@ieee.org, yuntao.wang@xjtu.edu.cn).}
\thanks{W. Chen and Q. Wu are with the Department of Electronic Engineering, Shanghai Jiao Tong University, Shanghai 200240, China (e-mail: wenchen@sjtu.edu.cn, qingqingwu@sjtu.edu.cn). }
\thanks{(\emph{Corresponding author: Zhou Su})}}



\maketitle

\begin{abstract}
In this paper, we investigate a secure communication architecture based on unmanned aerial vehicle (UAV), which enhances the security performance of the communication system through UAV trajectory optimization. We formulate a control problem of minimizing the UAV flight path and power consumption while maximizing secure communication rate over infinite horizon by jointly optimizing UAV trajectory, transmit beamforming vector, and artificial noise (AN) vector. Given the non-uniqueness of optimization objective and significant coupling of the optimization variables, the problem is a non-convex optimization problem which is difficult to solve directly. To address this complex issue, an alternating-iteration technique is employed to decouple the optimization variables. Specifically, the problem is divided into three subproblems, i.e., UAV trajectory, transmit beamforming vector, and AN vector, which are solved alternately. Additionally, considering the susceptibility of UAV trajectory to disturbances, the model predictive control (MPC) approach is applied to obtain UAV trajectory and enhance the system robustness. Numerical results demonstrate the superiority of the proposed optimization algorithm in maintaining accurate UAV trajectory and high secure communication rate compared with other benchmark schemes.
\end{abstract}

\begin{IEEEkeywords}
Secure communication, UAV trajectory optimization, model predictive control, system robustness.
\end{IEEEkeywords}

\section{Introduction}
\IEEEPARstart{I}{n} recent years, unmanned aerial vehicles (UAVs), commonly referred to as drones, have garnered extensive application in domains such as road monitoring, emergency response, 5G millimeter-wave radar communications, and wireless sensor networks due to their high maneuverability, ease of deployment, and line-of-sight communication capabilities \cite{7317490,4,8856195}. Despite the substantial advantages that UAVs offer in wireless communications, such as flexible deployment and efficient real-time communication capabilities, their inherent openness, sensitivity to disturbances, and the intrinsic nature of wireless communication pose significant security challenges for UAV-based communications. The open nature of UAVs renders their communication links vulnerable to unauthorized access and attacks, while disturbances can adversely affect signal stability and reliability \cite{6}. Moreover, the inherent openness of wireless communication further exacerbates the risk of signal interception and tampering \cite{7}. Consequently, research into the security of UAV-based communication systems has become critically important. Currently, in UAV secure communication networks, UAVs are usually employed as base stations (BSs) or relays to enhance network coverage and communication capabilities.

The utilization of UAVs as BSs to provide wireless communication services in special scenarios has become quite mature, and they can offer high-quality secure services to users in remote areas without ground BS coverage or for emergency communications \cite{11,12,13,14,15,16,17,18,19}. In \cite{11}, a cyclical iterative algorithm based on approximate beamforming patterns was developed to enhance the communication rates of users, thereby providing reliable communication for users stranded at sea. In \cite{12}, authors studied how to fully realize the potential of UAV-based technologies and applications in the face of several security challenges. In \cite{13}, authors introduced a secure downlink multi-user transmission scheme enabled by a flexible UAV-BS and non-orthogonal multiple access (NOMA). In \cite{14,15}, authors studied the physical layer security of a UAV network, where a UAV-BS transmitted confidential information to receivers in the presence of multiple eavesdroppers. In \cite{16}, a single-antenna multi-user communication system was proposed, where UAV-BS employed NOMA techniques to provide services to ground users. In \cite{17}, authors explored UAV-to-UAV (U2U) communication scenarios, applying deep Q-networks and actor-critic algorithms to maximize secure communication rates and smoothness. In \cite{18}, the authors investigated the secrecy fairness optimization in UAV-enabled wireless networks, utilizing cooperative rate-splitting to protect the downlink in a two-user multi-input single-output (MISO) system. In \cite{19}, authors investigated the effective network secrecy throughput of the uplink UAV network, in which they employed rate splitting multiple access (RSMA) for secure transmission of legitimate users.

Furthermore, the deployment of UAVs as relays within secure communication networks has also reached a significant level of sophistication. In scenarios requiring network coverage extension or enhancement of communication links, UAV relays effectively serve as intermediary nodes, bridging the primary communication network with user terminals. This application is particularly advantageous in areas with insufficient ground station coverage or dynamic environments, such as disaster relief operations or military missions. By employing UAV relays, one can substantially improve communication quality and stability, ensuring reliable data transmission and communication services even under the threat of eavesdroppers \cite{21,22,23,9798882,24}. In \cite{21}, UAVs served as relays to enhance the security of satellite-to-vehicle links, while also acting as jammers to generate artificial noise (AN) to confound potential eavesdroppers. In \cite{22}, the authors investigated the issue of secure transmission in a cache-enabled UAV-relaying network with device-to-device (D2D) communications in the presence of an eavesdropper. In \cite{23}, the authors proposed an artificial noise beamforming allocation scheme to combat eavedroppers for a full duplex UAV relaying scenario. In \cite{24}, authors proposed a secure short-packet communication system, where a UAV works by periodically receiving and relaying small data packets to its receive, in the presence of an eavesdropper. In \cite{26}, a secrecy outage performance achieved by opportunistic relaying for a low-altitude UAV swarm secure communication system was investigated in the face of multiple UAV-eavesdroppers.

In traditional research on UAV-assisted communications, offline algorithms have predominantly been utilized to address problems \cite{28,41,42}. However, when scenarios change or disturbances occur, UAV-assisted communication systems may struggle to achieve the desired performance. Recently, the model predictive control (MPC) has gained significant traction in the field of energy network optimization due to its real-time capabilities and robustness \cite{30,31,32,51,52,53,54}. The real-time and robust nature of MPC effectively mitigates UAV sensitivity to disturbances and assists in adjusting UAV trajectory based on real-time information, thereby enhancing system robustness. This makes MPC particularly well-suited for UAV-based secure communication systems. Consequently, in this paper, we implement the MPC to solve UAV trajectory and secure resource allocation, aiming to optimize system performance and address disturbances. 
MPC has been extensively studied as a technique that optimizes control performance under uncertainty in the domain of energy network optimization and scheduling. As demonstrated in \cite{30}, the authors employed a robust nonlinear model predictive control with generalized regional tracking to optimize an energy facility integrating battery storage and solar photovoltaic systems for hydrogen production. In \cite{31}, a risk-averse MPC scheme was proposed for managing energy networks with a high proportion of renewable energy sources, mitigating the potential risks associated with renewable energy uncertainties. In \cite{32}, the authors introduced an energy management framework based on the Tube-based MPC that sacrifices minimal economic gains and computational efficiency to enhance the robustness of energy dispatch strategies against system uncertainties. Recently, applying MPC algorithmic principles to enhance the robustness of UAV communication models has garnered significant attention. In \cite{33}, the authors proposed a predictive control model at each time step based on the statistical information of known A2G and ground-to-air (G2A) wireless channels, thereby enhancing the system's robustness to statistical information uncertainties. 

In the paradigm of UAV-assisted communication networks, the imperatives of system security and robustness are becoming increasingly pronounced, catalyzing a surge in research endeavors focused on developing UAV-based secure communication systems. However, studies addressing communication security and system robustness remain relatively scarce, warranting further investigation and analysis. Thus, leveraging the real-time and robust characteristics of MPC effectively addresses the imprecision in UAV positioning caused by disturbances, thereby significantly enhancing the performance of UAV-based communication systems. In this paper, we explore the problem of navigating a UAV from a specified starting point to a designated endpoint, while considering factors such as communication security for users, UAV power consumption, and AN, etc. The main contributions of this paper are as follows:

\begin{itemize}
\item{We formulate a control problem over an infinite time horizon to describe the process of a UAV navigating from a given starting point to a designated endpoint while providing secure communication services to users. Subsequently, considering the sensitivity of UAV-assisted communication to disturbances and the limitations of offline algorithms, we apply MPC to refine the formulated problem. This enhancement enables the UAV to make real-time adjustments for position deviations and other uncertainties, thereby improving the system's robustness.}
\item{The refined optimization control problem, characterized by time-domain rolling, involves multiple optimization objectives and exhibits non-convex characteristics. To address this, we first apply alternating-iteration technique to decouple the optimization problem into three subproblems, i.e., the UAV trajectory, transmit beamforming vector, and AN vector. Then, internal convex approximation technique is used to address the non-convexity of objective functions and constraints. By iteratively solving the abovementioned subproblems, the UAV trajectory optimization and secure resource allocation scheme can be obtained.}
\item{Numerical results validate the superiority of the proposed algorithm in terms of performance advantage compared with other benchmarks. Compared to the conventional block coordinate descent (BCD) algorithm, UAVs equipped with MPC demonstrate the capability to rectify positional discrepancies in real-time under perturbation, thereby ensuring precise navigation to the designated endpoint while maintaining a stable communication link for legitimate users, with reduced energy consumption. Moreover, simulation results validate that the proposed algorithm robustly facilitates the UAV’s auxiliary communication duties across varying levels of perturbations, substantially enhancing the robustness of the communication system.}
\end{itemize}

\textit{Notations:} In this paper, scalars are denoted by lower-case letters. Bold lowercase and uppercase letters are used to denote vectors and matrices, respectively. $\mathbb{R}{^{m \times n}}$ and $\mathbb{C}{^{m \times n}}$ represent the $m \times n$ dimensional real and complex fields, respectively. The terms $|\cdot|$ and $\|\cdot\|$ denote the absolute value of a complex scalar and the Euclidean norm of a vector, respectively. ${{\bf{{I}}}_N}$ signifies an $N$-dimensional identity matrix. For a square matrix $\bf{A}$, ${{\bf{A}}^T}$ and ${{\bf{A}}^H}$ respectively indicate the transpose and conjugate transpose of matrix ${\bf{{A}}}$, while ${\rm{Tr}}\left( {\bf{A}} \right)$ and ${\rm{Rank}}\left( {\bf{{A}}} \right)$ represent the trace and rank of matrix ${\bf{{A}}}$, respectively. $j$ is the imaginary unit, i.e., $j^2 = -1$. $\mathbb{E}\left\{  \cdot  \right\}$ is expectation operator. $ \sim $ denotes "is distributed as", and the distribution of a circularly symmetric complex Gaussian (CSCG) random vector with mean $\mu$ and covariance matrix $\bf{C}$ can be expressed as $ {\cal C}{\cal N}\left( {\mu,\bf{C}} \right)$. Finally, ${\nabla _{\bf{x}}}f\left( {\bf{x}} \right)$ represents the gradient vector of function $f\left( {\bf{x}} \right)$ with respect to ${\bf{x}}$.

\section{UAV-Enabled Secure Communication System Model and Problem Formulation}
\subsection{Network Model}
As depicted in Fig. 1, this paper considers a UAV-enabled multi-user secure communication network, comprising a UAV equipped with a dual-functional radar and communication (DFRC) BS (hereinafter referred to as UAV), $K$ legitimate users, and a potential eavesdropper. The UAV is outfitted with a uniform linear array (ULA) composed of $N$ antennas, while the $K$ legitimate users and the potential eavesdropper are equipped with single antenna. The set of legitimate users is denoted by $\cal K$. This paper considers a two-phase operational cycle, specifically the UAV scanning phase $T_{\textrm{scan}}$ and the UAV secure communication phase $T_{\textrm{secu}}$. During the UAV scanning phase $T_{\textrm{scan}}$, the UAV transmits a detection signal to detect the suspicious target. Utilizing existing parameter estimation methodologies, such as those delineated in \cite{10403776,10634583}, it determines the angle and distance of the suspicious target based on the echo signals, thereby ascertaining its location. It is noteworthy that the suspicious target detected during the UAV scanning phase $T_{\textrm{scan}}$ is presumed to be the potential eavesdropper for the subsequent UAV secure communication phase $T_{\textrm{secu}}$. In the UAV secure communication phase $T_{\textrm{secu}}$, the information about the eavesdropper obtained earlier is utilized to enhance the security communication performance of by designing the UAV trajectory and transmit beamforming. Moreover, to augment the security performance during this phase, beams carrying information alongside AN beams are simultaneously transmitted from the UAV.

\begin{figure}[t]
	\centering  
	\includegraphics[scale=0.30]{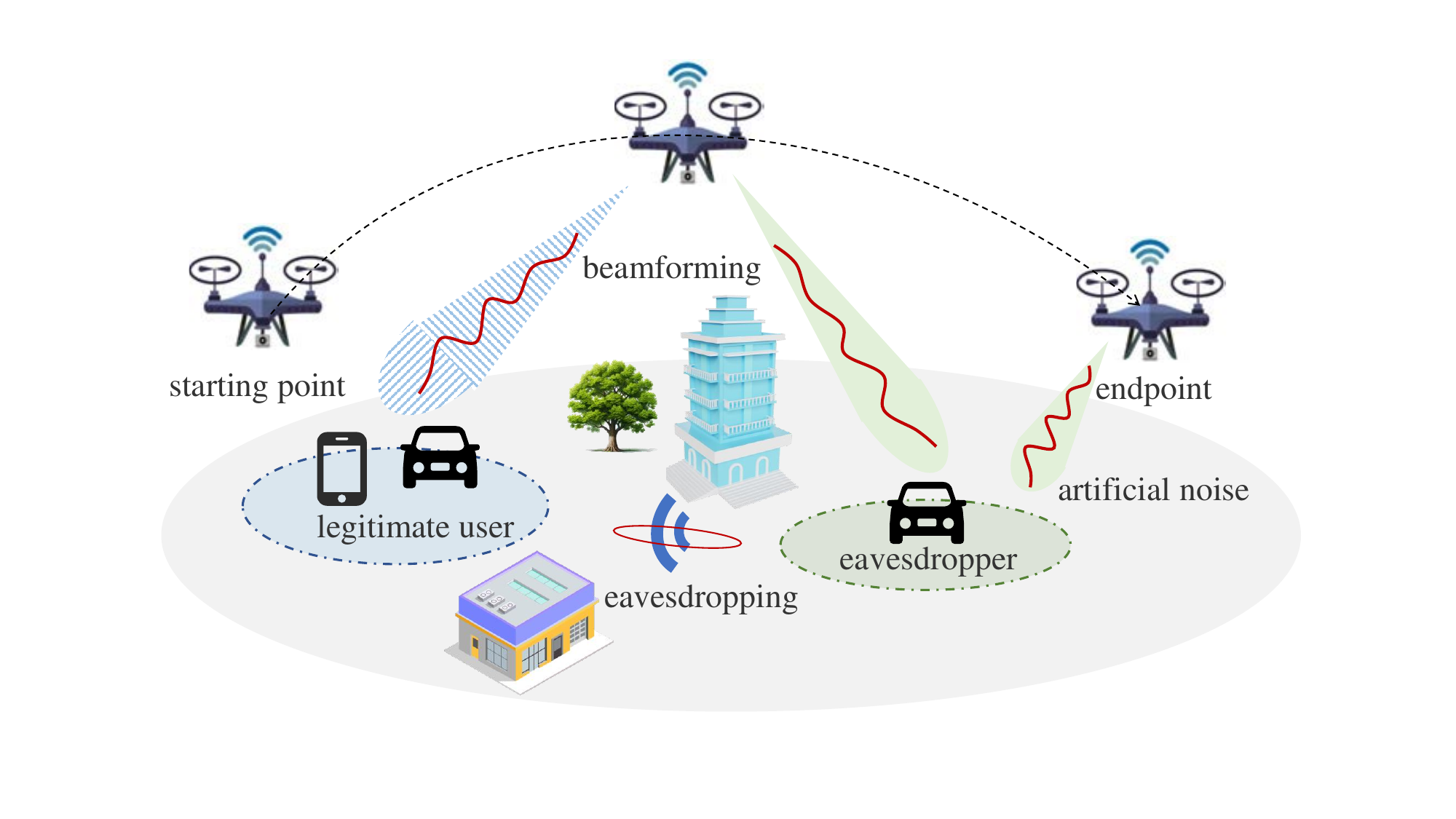}
	\caption{UAV-enabled secure communication networks.}  
	\label{Figure 1} 
\end{figure}


\subsection{UAV Motion Model and Channel Model}
We consider a UAV equipped with a fixed-wing, navigating from a designated starting point ${{\bf{q}}_A} = {\left( {{x_A},{y_A},{z_A}} \right)^T} \in {\mathbb{R}^3}$ to a specified endpoint ${{\bf{q}}_B} = {\left( {{x_B},{y_B},{z_B}} \right)^T} \in {\mathbb{R}^3}$. The flight is segmented into $I$ time slots, i.e., $i=0,1,...,I-1$, during which the three-dimensional dynamic model of the UAV can be modeled as follows
\begin{equation}
    {\bf{q}}\left( {\rm{0}} \right) = {{\bf{q}}_A},
\end{equation}
\begin{equation}
    {\bf{q}}\left( {i + 1} \right) = {\bf{q}}\left( i \right) + {\bf{v}}\left( i \right){t_c},
\end{equation}
\begin{equation}
    {Z_{\min }} \le {q_z}\left( i \right) \le {Z_{\max }},
\end{equation}
\begin{equation}
    \left\| {\left( {{v_x}\left( i \right),{v_y}\left( i \right)} \right)} \right\| \le {V_{\max }},
\end{equation}
\begin{equation}
    {v_z}\left( i \right) \le {U_{\max }},
\end{equation}
\begin{equation}
    \left\| {{\bf{v}}\left( {i + 1} \right) - {\bf{v}}\left( i \right)} \right\| \le {{a}_{\max }}{t_c},
\end{equation}
where ${\bf{q}}\left( i \right)  = {\left( {{q_x}\left( i \right),{q_y}\left( i \right),{q_z}\left( i \right)} \right)^T}$ and ${\bf{v}}\left( i \right)  = {\left( {{v_x}\left( i \right),{v_y}\left( i \right),{v_z}\left( i \right)} \right)^T}$ respectively denote the position and velocity vectors of the UAV at the $i$-th time slot, while ${Z_{{\rm{min}}}}$ and ${Z_{{\rm{max}}}}$ represent the minimum and maximum flight altitudes of the UAV. The maximum values of the UAV's horizontal speeds ${{\bf{v}}_h}\left( i \right) = {\left( {{v_x}\left( i \right),{v_y}\left( i \right)} \right)^T}$ and vertical speeds ${v_z}\left( i \right)$ are represented by ${V_{{\rm{max}}}}$ and ${U_{{\rm{max}}}}$, respectively. Additionally, ${a_{\max }}$ denotes the maximum acceleration of the UAV, and ${t_c}$ represents the time interval of each slot.

During the $i$-th time slot, the power consumption of the UAV can be expressed as \cite{34}
\begin{equation}
    \begin{aligned}
            P\left( {{\bf{v}}\left( i \right)} \right) &= \frac{{{W^2}}}{{\sqrt 2 \rho S}}{\left( {{{\left\| {{{\bf{v}}_h}\left( i \right)} \right\|}^2} + \sqrt {{{\left\| {{{\bf{v}}_h}\left( i \right)} \right\|}^4} + 4V_h^4} } \right)^{ - \frac{1}{2}}} \\
            & + W{v_z}\left( i \right) + \frac{{\zeta \rho S}}{8}{\left\| {{{\bf{v}}_h}\left( i \right)} \right\|^3},
    \end{aligned}
\end{equation}
where $W$ represents the mass of the UAV, $\rho $ denotes the air density, $S$ signifies the rotor disk area of the UAV, ${V_h} = \sqrt {\frac{W}{{2\rho S}}} $ is the hover parameter, and $\zeta $ is the UAV's profile drag coefficient. The first term in Eq. (7) corresponds to the power consumption of the UAV during horizontal flight, the second term accounts for the power consumption during vertical lift, and the third term represents the blade drag power consumption of the UAV.

Furthermore, we discuss the channel model for the UAV-enabled secure communication system. As the UAV navigates from the starting point ${{\bf{q}}_A}$ to the endpoint ${{\bf{q}}_B}$, it provides secure communication services to legitimate users. At the $i$-th time slot, for the $k$-th legitimate user located at coordinates ${{\bf{q}}_k} = {\left( {{x_k},{y_k},0} \right)^T}$, the channel between the UAV and the legitimate user ${\bf{\bar h}}_k^H\left( i \right) \in {\mathbb{C}{^{1 \times N}}} $ is modeled as a line-of-sight (LoS) channel, which can be expressed as follows
\begin{equation}
    {\bf{\bar h}}_k^H\left( i \right) = \left[ {1,{e^{ - j2\pi \frac{d}{\lambda }\sin \theta \left( i \right)}},...,{e^{ - j2\pi \left( {N - 1} \right)\frac{d}{\lambda }\sin \theta \left( i \right)}}} \right],
\end{equation}
where $\lambda $ denotes the wavelength, and $d$ represents the spacing between antenna elements, typically set to ${\lambda  \mathord{\left/{\vphantom {\lambda  2}} \right. \kern-\nulldelimiterspace} 2}$. The term $\theta \left( i \right)$ corresponds to the angle-of-departure (AoD) at the UAV antenna during the $i$-th time slot. Consequently, the channel gain between the UAV and the $k$-th legitimate user during the $i$-th time slot can be expressed as
\begin{equation}
    {\bf{h}}_k^H\left( i \right) = \sqrt {{g_0}{{\left\| {{\bf{q}}\left( i \right) - {{\bf{q}}_k}} \right\|}^{ - \alpha_1 }}} {\bf{\bar h}}_k^H\left( i \right),
\end{equation}
where ${g_0}$ represents the path loss at a reference distance of one meter, and $\alpha _1 $ denotes the path loss exponent.

Similarly, the channel between the UAV and the potential eavesdropper ${\bf{\bar g}}_e^H\left( i \right) \in {\mathbb{C}{^{1 \times N}}} $ is also modeled as a LoS channel, denoted by
\begin{equation}
    {\bf{\bar g}}_e^H\left( i \right) = \left[ {1,{e^{ - j2\pi \frac{d}{\lambda }\sin \varphi \left( i \right)}},...,{e^{ - j2\pi \left( {N - 1} \right)\frac{d}{\lambda }\sin \varphi \left( i \right)}}} \right],
\end{equation}
where $\varphi (i)$ denotes the AoD at the UAV antenna for the potential eavesdropper in the $i$-th time slot. Consequently, the channel gain between the UAV and the potential eavesdropper during the $i$-th time slot can be expressed as follows
\begin{equation}
    {\bf{g}}_e^H\left( i \right) = \sqrt {{g_0}{{\left\| {{\bf{q}}\left( i \right) - {{\bf{q}}_e}} \right\|}^{ - \alpha_2 }}} {\bf{\bar g}}_e^H\left( i \right).
\end{equation}
The AoD at the UAV antenna $\varphi \left( i \right)$ and the position of the potential eavesdropper ${{\bf{q}}_e}$ in the $i$-th time slot can be acquired through the scanning phase ${T_\textrm{scan}}$ of the UAV.
\newcounter{my1}
\begin{figure*}[!t]
	\normalsize
	\setcounter{my1}{\value{equation}}
	\setcounter{equation}{15}
	\begin{equation}
		\begin{aligned}
			{R_k}\left( i \right) = {\log _2}\left( {1 + \frac{{{{\left| {{\bf{h}}_k^H\left( i \right){{\bf{w}}_k}\left( i \right)} \right|}^2}}}{{\sum\limits_{r \in {\cal K}\backslash \left\{ k \right\}} {{{\left| {{\bf{h}}_k^H\left( i \right){{\bf{w}}_r}\left( i \right)} \right|}^2} + {{\left| {{\bf{h}}_k^H\left( i \right){\bf{m}}\left( i \right)} \right|}^2} + \sigma _k^2\left( i \right)} }}} \right).
		\end{aligned}
	\end{equation}
	\setcounter{equation}{\value{my1}}
\hrulefill
\vspace*{4pt}
\end{figure*}
\subsection{Signal Model and Performance Metric}
In the $i$-th time slot, the UAV provides secure communication services to $K$ legitimate users. To enhance the performance of secure communications, an AN signal is incorporated into the data stream. Consequently, the transmission signal of the UAV in the $i$-th time slot can be represented as
\begin{equation}
    {\bf{x}}\left( i \right) = \sum\limits_{k \in {\cal K}} {{{\bf{w}}_k}\left( i \right)} {b_k}\left( i \right) + {\bf{m}}\left( i \right),
\end{equation}
where ${{\bf{w}}_k}\left( i \right) \in {\mathbb{C}{^{N \times 1}}}$ denotes the beamforming vector allocated to the $k$-th legitimate user, and ${b_k}\left( i \right) \in {\mathbb{C}}$ represents the signal transmitted to the $k$-th legitimate user, assumed to satisfy ${\mathbb{E}}\left\{ {{{\left| {{b_k}\left( i \right)} \right|}^2}} \right\} = 1$. ${\bf{m}}\left( i \right) \in {\mathbb{C}{^{N \times 1}}}$ represents the AN signal transmitted from the UAV, which follows CSCG distribution, expressed as follows
\begin{equation}
    {\bf{m}}\left( i \right) \sim {\cal C}{\cal N}\left( {{\bf{0}},{\bf{M}}\left( i \right)} \right),
\end{equation}
where ${\bf{M}}\left( i \right) \succeq 0,{\bf{M}}\left( i \right) \in {\mathbb{H}{^{N}}}$ denotes the covariance matrix of the AN signal, and $\mathbb{H}$ denotes the Hermitian matrix. Thus, in the $i$-th time slot, the signal received by the $k$-th legitimate user can be expressed as
\begin{equation}
    \begin{aligned}
           {y_k}\left( i \right) &= \underbrace {{\bf{h}}_k^H\left( i \right){{\bf{w}}_k}\left( i \right){b_k}\left( i \right)}_{{\textrm{Desired~signal}}} + \underbrace {\sum\limits_{r \in {\cal K}\backslash \left\{ k \right\}} {{\bf{h}}_k^H\left( i \right){{\bf{w}}_r}\left( i \right)} {b_r}\left( i \right)}_{{\textrm{Multiuser~interference}}} \\
            & +\underbrace {{\bf{h}}_k^H\left( i \right){\bf{m}}\left( i \right)}_{{\textrm{Artificial~noise}}} + \underbrace {{n_k}\left( i \right)}_{{\textrm{Noise}}},
    \end{aligned}
\end{equation}
where ${n_k}\left( i \right) \sim {\cal C}{\cal N}\left( {0,\sigma _k^2\left( i \right)} \right)$ denotes the additive white Gaussian noise (AWGN) introduced at the receiving antenna of the $k$-th legitimate user in the $i$-th time slot. Similarly, the signal received by the potential eavesdropper in the $i$-th time slot can be expressed as
\begin{equation}
    {y_e}\left( i \right) = {\bf{g}}_e^H\left( i \right)\sum\limits_{k \in {\cal K}} {{{\bf{w}}_k}\left( i \right)} {b_k}\left( i \right) + {\bf{g}}_e^H\left( i \right){\bf{m}}\left( i \right) + {n_e}\left( i \right),
\end{equation}
where ${n_e}\left( i \right) \sim {\cal C}{\cal N}\left( {0,\sigma _e^2\left( i \right)} \right)$ denotes the AWGN introduced at the receiving antenna of the potential eavesdropper in the $i$-th time slot.

Consequently, according to the Shannon formula, the achievable rate (bps/Hz) for the $k$-th legitimate user in the $i$-th time slot can be expressed as Eq. (16). To ensure the security of communication, this paper considers the worst-case scenario wherein potential eavesdropper is capable of decoding the communication information of legitimate users, unaffected by multi-user interference. Therefore, in the $i$-th time slot, the eavesdropping rate (bps/Hz) for the $k$-th legitimate user by the potential eavesdropper can be expressed as
\setcounter{equation}{16}
\begin{equation}
    {C_{e,k}}\left( i \right) = {\log _2}\left( {1 + \frac{{{{\left| {{\bf{g}}_e^H\left( i \right){{\bf{w}}_k}\left( i \right)} \right|}^2}}}{{{{\left| {{\bf{g}}_e^H\left( i \right){\bf{m}}\left( i \right)} \right|}^2} + \sigma _e^2\left( i \right)}}} \right).
\end{equation}
Thus, in the $i$-th time slot, the maximum secure communication rate of the $k$-th legitimate user can be expressed as
\begin{equation}
   R_k^s\left( i \right) = {\left[ {{R_k}\left( i \right) - {C_{e,k}}\left( i \right)} \right]^ + }, 
\end{equation}
where ${\left[  \cdot  \right]^ + } = \max \left\{ {0, \cdot } \right\}$.

\subsection{Problem Formulation Based on Model Predictive Control}
In this paper, we formulate a control problem defined over an infinite time horizon, where an UAV is required to navigate from a specified starting point ${{\bf{q}}_A}$ to an endpoint ${{\bf{q}}_B}$. Under constraints of finite power consumption, user quality of service, and UAV motion limitations, our objective is to minimize UAV flight path and power consumption while maximize secure communication rate of legitimate users. This problem can be expressed as problem (P1) as follows
\begin{subequations}
	\begin{align}
		{\textrm{(P1)}}~~~~&\mathop {\min }\limits_{\mathop {{\bf{q}}\left( i \right),{\bf{v}}\left( i \right),{{\bf{w}}_k}\left( i \right),{\bf{m}}\left( i \right)}\limits_{i = 0,1, \ldots } } \sum {{\varpi _1}} {\left\| {{\bf{q}}\left( i \right) - {{\bf{q}}_B}} \right\|^2} + {\varpi _2}P\left( {{\bf{v}}\left( i \right)} \right) \nonumber \\
        &- {\varpi _3}\sum\limits_{k \in {\cal K}} {R_k^s\left( i \right)},\nonumber \\
		{\textrm{s.t.}}\qquad &{\bf{q}}\left( {i + 1} \right) = {\bf{q}}\left( i \right) + {\bf{v}}\left( i \right){t_c},\\
		&\left\| {\left( {{v_x}\left( i \right),{v_y}\left( i \right)} \right)} \right\| \le {V_{\max }},\\
		&\left| {{v_z}\left( i \right)} \right| \le {U_{\max }},\\
		&\left\| {{\bf{v}}\left( {i + 1} \right) - {\bf{v}}\left( i \right)} \right\| \le {a_{\max }}{t_c},\\
		&{Z_{\min }} \le {{q}_z}\left( i \right) \le {Z_{\max }},\\
		&\sum\limits_{k = 1}^K {{{\left\| {{{\bf{w}}_k}\left( i \right)} \right\|}^2}} \! +\! {\left\| {{\bf{m}}\left( i \right)} \right\|^2} \!+\! P\left( {{\bf{v}}\left( i \right)} \right) \le {P_{\max }},\\
		&{R_k}\left( i \right) \ge {R_{\min }},\\
		&{C_{e,k}}\left( i \right) \le {R_{\max }},
	\end{align}
\end{subequations}
where the objective function consists of three terms: the first aims to optimize the UAV's flight path from the starting point to the destination at the highest possible speed. The second seeks to minimize power consumption. The third ensures secure communications for legitimate users, with ${\varpi _1},{\varpi _2},{\varpi _3}>0$ serving as a constant that can be adjusted to prioritize these objectives. Constraint (19a) defines the recursive relationship between two consecutive time instances. Constraints (19b) and (19c) limit the UAV's horizontal and vertical velocities, respectively. Constraint (19d) prevents abrupt changes in velocity to ensure safety. Constraint (19e) maintains the UAV's flight altitude within an optimal range for visibility. Constraint (19f) imposes power limitations on the UAV. Constraint (19g) sets a minimum communication rate for users to enhance quality of service. Constraint (19h) limits the maximum eavesdropping rate of a potential eavesdropper to ensure transmission security. At this juncture, we have successfully established a model for the UAV-based secure communication system within an infinite time domain. Nevertheless, in order to enhance the robustness of the system further, we employ the MPC algorithm, thereby effecting the following optimizations and refinements.

Let ${N_p}$ be the prediction step and ${\cal I}  = \left\{ {i,i + 1, \ldots ,i + {N_p}} \right\}$. In the $i$-th time slot, the optimal solution for $\left( {{\bf{q}}\left( i \right),{\bf{v}}\left( i \right),{{\bf{w}}_k}\left( i \right),{\bf{m}}\left( i \right)} \right)$ over the rolling horizon $\left[ {i,i + {N_p}} \right]$ can be determined by solving the MPC problem (P2) as follows
\begin{subequations}
	\begin{align}
		{\textrm{(P2)}}~~~~&\mathop {\min }\limits_{\mathop {{\bf{q}}\left( {i'} \right),{\bf{v}}\left( {i'} \right),{{\bf{w}}_k}\left( {i'} \right),{\bf{m}}\left( {i'} \right)} } \sum\limits_{i' = i}^{i + {N_p}} {{\varpi _1}} {\left\| {{\bf{q}}\left( {i'} \right) - {{\bf{q}}_B}} \right\|^2}  \nonumber \\
        &+ {\varpi _2}P\left( {{\bf{v}}\left( {i'} \right)} \right)  - {\varpi _3}\sum\limits_{k \in K} {R_k^s\left( {i'} \right)},\nonumber \\
		{\textrm{s.t.}}\qquad &{\bf{q}}\left( {i' + 1} \right) = {\bf{q}}\left( {i'} \right) + {\bf{v}}\left( {i'} \right){t_c},i' \in {\cal I}\backslash \left\{ {i + {N_p}} \right\},\\
		&\left\| {\left( {{v_x}\left( {i'} \right),{v_y}\left( {i'} \right)} \right)} \right\| \le {V_{\max }},\\
		&\left| {{v_z}\left( {i'} \right)} \right| \le {U_{\max }},\\
		&\left\| {{\bf{v}}\left( {i' + 1} \right) - {\bf{v}}\left( {i'} \right)} \right\| \le {a_{\max }}{t_c},i' \in {\cal I}\backslash \left\{ {i + {N_p}} \right\},\\
		&{Z_{\min }} \le {q_z}\left( {i'} \right) \le {Z_{\max }},\\
		&\sum\limits_{k = 1}^K {{{\left\| {{{\bf{w}}_k}\left( {i'} \right)} \right\|}^2}} \! + \! {\left\| {{\bf{m}}\left( {i'} \right)} \right\|^2} \! + \! P\left( {{\bf{v}}\left( {i'} \right)} \right) \le {P_{\max }},\\
		&{R_k}\left( {i'} \right) \ge {R_{\min }},\\
		&{C_{e,k}}\left( {i'} \right) \le {R_{\max }}.
	\end{align}
\end{subequations}
The problem (P1) is transformed from an optimization problem in an infinite time domain to a control problem (P2) based on MPC. Nevertheless, the problem’s inherent highly coupling and non-convex nature render direct solution approaches exceptionally challenging. Consequently, this paper employs an alternating-iteration and internal convex approximation techniques to reformulate the problem, with the detailed procedures to be delineated in the subsequent section.
\section{Joint UAV Trajectory Optimization and Secure Resource Allocation Based on MPC}
\subsection{Problem Transformation}
Given the presence of non-convex functions in the aforementioned problem (P2), this paper facilitates matrix lifting technique to transform the optimization problem (P2). Let ${\bf{W}}_k(i') = {\bf{w}}_k(i'){\bf{w}}_k^H(i')$, the transformed problem (P3) can be expressed as follows
\begin{subequations}
	\begin{align}
		{\textrm{(P3)}}~~~~&\mathop {\min }\limits_{\mathop {{\bf{q}}\left( {i'} \right),{\bf{v}}\left( {i'} \right),{{\bf{W}}_k}\left( {i'} \right),{\bf{M}}\left( {i'} \right)} } \sum\limits_{i' = i}^{i + {N_p}} {{\varpi _1}} {\left\| {{\bf{q}}\left( {i'} \right) - {{\bf{q}}_B}} \right\|^2}  \nonumber \\
        &+ {\varpi _2}P\left( {{\bf{v}}\left( {i'} \right)} \right)  - {\varpi _3}\sum\limits_{k \in K} {R_k^s\left( {i'} \right)},\nonumber \\
		{\textrm{s.t.}}\qquad &{\bf{q}}\left( {i' + 1} \right) = {\bf{q}}\left( {i'} \right) + {\bf{v}}\left( {i'} \right){t_c},i' \in {\cal I}\backslash \left\{ {i + {N_p}} \right\},\\
		&\left\| {\left( {{v_x}\left( {i'} \right),{v_y}\left( {i'} \right)} \right)} \right\| \le {V_{\max }},\\
		&\left| {{v_z}\left( {i'} \right)} \right| \le {U_{\max }},\\
		&\left\| {{\bf{v}}\left( {i' + 1} \right) - {\bf{v}}\left( {i'} \right)} \right\| \le {a_{\max }}{t_c},i' \in {\cal I}\backslash \left\{ {i + {N_p}} \right\},\\
		&{Z_{\min }} \le {q_z}\left( {i'} \right) \le {Z_{\max }},\\
		&\sum\limits_{k = 1}^K {{\rm{Tr}}\left( {{{\bf{W}}_k}\left( {i'} \right)} \right)}\!  + \!{\rm{Tr}}\left( {{\bf{M}}\left( {i'} \right)} \right) \!+ \!P\left( {{\bf{v}}\left( {i'} \right)} \right) \le {P_{\max }},\\
        &{R_k}\left( {i'} \right) \ge {R_{\min }},\\
		&{C_{e,k}}\left( {i'} \right) \le {R_{\max }},\\
  	&{{\bf{W}}_k}\left( {i'} \right) \succeq 0,\\
		&{\rm{Rank}}\left( {{{\bf{W}}_k}\left( {i'} \right)} \right) = 1,\\
        &{\bf{M}}\left( {i'} \right) \succeq 0,\\
        &{\rm{Rank}}\left( {{\bf{M}}\left( {i'} \right)} \right) = 1,
        \end{align}
\end{subequations}
where constraints (21i)-(21l) are semi-positive definite and rank-1 constraints introduced after matrix lifting of the transmit beamforming vector and the AN vector. However, due to the highly coupling of multiple optimization objectives within the objective function, direct resolution is challenging. Consequently, the problem (P3) is decomposed into three subproblems as follows, which are solved iteratively through an alternating-iteration approach. To enhance the robustness of the system, this paper employs the MPC algorithm for effective problem-solving.

\newcounter{my2}
\begin{figure*}[!t]
	\normalsize
	\setcounter{my2}{\value{equation}}
	\setcounter{equation}{35}
	\begin{equation}
            \begin{aligned}
                {R_k}\left( {i'} \right) &= {{\log }_2}\left( {1 + \frac{{{\rm{Tr}}\left( {{{\bf{W}}_k}\left( {i'} \right)} \right){g_0}{{\left\| {{\bf{q}}\left( {i'} \right) - {{\bf{q}}_k}} \right\|}^{ - {\alpha _1}}}N}}{{{\sum\limits _{r \in {\cal K}\backslash \{ k\} }}{\rm{Tr}}\left( {{{\bf{W}}_r}\left( {i'} \right)} \right){g_0}{{\left\| {{\bf{q}}\left( {i'} \right) - {{\bf{q}}_r}} \right\|}^{ - {\alpha _1}}}N + {\rm{Tr}}\left( {{\bf{M}}\left( {i'} \right)} \right){g_0}{{\left\| {{\bf{q}}\left( {i'} \right) - {{\bf{q}}_k}} \right\|}^{ - {\alpha _1}}}N + \sigma _k^2\left( {i'} \right)}}} \right)\\
                & = \underbrace {{{\log }_2}\left( {\sum\limits_{r \in {\cal K}} {{\rm{Tr}}\left( {{{\bf{W}}_r}\left( {i'} \right)} \right){g_0}{{\left\| {{\bf{q}}\left( {i'} \right) - {{\bf{q}}_r}} \right\|}^{ - {\alpha _1}}}N}  + {\rm{Tr}}\left( {{\bf{M}}\left( {i'} \right)} \right){g_0}{{\left\| {{\bf{q}}\left( {i'} \right) - {{\bf{q}}_k}} \right\|}^{ - {\alpha _1}}}N + \sigma _k^2\left( {i'} \right)} \right)}_{R_k^{\rm{1}}}\\
                & - \underbrace {{{\log }_2}\left( {{\sum _{r \in {\cal K}\backslash \{ k\} }}{\rm{Tr}}\left( {{{\bf{W}}_r}\left( {i'} \right)} \right){g_0}{{\left\| {{\bf{q}}\left( {i'} \right) - {{\bf{q}}_r}} \right\|}^{ - {\alpha _1}}}N + {\rm{Tr}}\left( {{\bf{M}}\left( {i'} \right)} \right){g_0}{{\left\| {{\bf{q}}\left( {i'} \right) - {{\bf{q}}_k}} \right\|}^{ - {\alpha _1}}}N + \sigma _k^2\left( {i'} \right)} \right)}_{R_k^{\rm{2} }},
            \end{aligned}
        \end{equation}
        \begin{equation}
            {\nabla _{{\bf{q}}\left( {i'} \right)}}R_k^2 \left( {{\bf{q}}\left( {i'} \right)} \right)\! = \!\frac{{ - {\alpha _1}{g_0}N\!\left( {\sum\limits_{r \in {\cal K}\backslash \{ k\} } {{{\left\| {{\bf{q}}\left( {i'} \right)\! -\! {{\bf{q}}_r}} \right\|}^{ - {\alpha _1}\! -\! 2}}{{\left( {{\bf{q}}\left( {i'} \right)\! -\! {{\bf{q}}_r}} \right)}^T}{\rm{Tr}}\left( {{{\bf{W}}_r}\left( {i'} \right)} \right)\! + \!{{\left\| {{\bf{q}}\left( {i'} \right) \!- \!{{\bf{q}}_k}} \right\|}^{ - {\alpha _1} \!-\! 2}}{{\left( {{\bf{q}}\left( {i'} \right)\! -\! {{\bf{q}}_k}} \right)}^T}{\rm{Tr}}\left( {{\bf{M}}\left( {i'} \right)} \right)} } \!\right)}}{{\ln 2\left( {\sum\limits_{r \in {\cal K}\backslash \{ k\} } {{\rm{Tr}}\left( {{{\bf{W}}_r}\left( {i'} \right)} \right){g_0}{{\left\| {{\bf{q}}\left( {i'} \right) - {{\bf{q}}_r}} \right\|}^{ - {\alpha _1}}}N + {\rm{Tr}}\left( {{\bf{M}}\left( {i'} \right)} \right){g_0}{{\left\| {{\bf{q}}\left( {i'} \right) - {{\bf{q}}_k}} \right\|}^{ - {\alpha _1}}}N + \sigma _k^2\left( {i'} \right)} } \right)}},
        \end{equation}
	\setcounter{equation}{\value{my2}}
\hrulefill
\vspace*{4pt}
\end{figure*}

\subsection{UAV Trajectory Optimization}
In optimizing the trajectory of UAV, transmit beamforming matrix and AN matrix are fixed that satisfy the constraints. Then, the UAV trajectory optimization problem (P4) is written as follows
\begin{subequations}
	\begin{align}
		{\textrm{(P4)}}~~~~&\mathop {\min }\limits_{\mathop {{\bf{q}}\left( {i'} \right),{\bf{v}}\left( {i'} \right)}} \sum\limits_{i' = i}^{i + {N_p}} {{\varpi _1}} {\left\| {{\bf{q}}\left( {i'} \right) - {{\bf{q}}_B}} \right\|^2} + {\varpi _2}P\left( {{\bf{v}}\left( {i'} \right)} \right)\nonumber\\
        &- {\varpi _3}\sum\limits_{k \in K} {R_k^s\left( {i'} \right)},\nonumber \\
		{\textrm{s.t.}}\qquad &{\bf{q}}\left( {i' + 1} \right) = {\bf{q}}\left( {i'} \right) + {\bf{v}}\left( {i'} \right){t_c},i' \in {\cal I}\backslash \left\{ {i + {N_p}} \right\},\\
		&\left\| {\left( {{v_x}\left( {i'} \right),{v_y}\left( {i'} \right)} \right)} \right\| \le {V_{\max }},\\
		&\left| {{v_z}\left( {i'} \right)} \right| \le {U_{\max }},\\
		&\left\| {{\bf{v}}\left( {i' + 1} \right) - {\bf{v}}\left( {i'} \right)} \right\| \le {a_{\max }}{t_c},i' \in {\cal I}\backslash \left\{ {i + {N_p}} \right\},\\
		&{Z_{\min }} \le {q_z}\left( {i'} \right) \le {Z_{\max }},\\
		&\sum\limits_{k = 1}^K {{\rm{Tr}}\left( {{{\bf{W}}_k}\left( {i'} \right)} \right)}\!  + \!{\rm{Tr}}\left( {{\bf{M}}\left( {i'} \right)} \right) \!+ \!P\left( {{\bf{v}}\left( {i'} \right)} \right) \le {P_{\max }},\\
        &{R_k}\left( {i'} \right) \ge {R_{\min }},\\
		&{C_{e,k}}\left( {i'} \right) \le {R_{\max }}.
	\end{align}
\end{subequations}
In the problem (P4), $P\left( {{\bf{v}}\left( {i'} \right)} \right)$, $R_k^s\left( {i'} \right)$, ${R_k}\left( {i'} \right)$ and ${C_{e,k}}\left( {i'} \right) $ exhibit non-convex characteristics, and we address them sequentially through the following procedural steps.

According to Eq. (7), the power consumption of the UAV during the $i'$-th time slot can be expressed as follows
\begin{equation}
    \begin{aligned}
        P\left( {{\bf{v}}\left( {i'} \right)} \right) &= \frac{{{W^2}}}{{\sqrt 2 \rho S}}{\left( {{{\left\| {{{\bf{v}}_h}\left( {i'} \right)} \right\|}^2} +\sqrt {{{\left\| {{{\bf{v}}_h}\left( {i'} \right)} \right\|}^4} + 4{V_h}^4} } \right)^{ - \frac{1}{2}}} \\
        &+ W{v_z}\left( {i'} \right) + \frac{{\zeta \rho S}}{8}{\left\| {{{\bf{v}}_h}\left( i' \right)} \right\|^3},
    \end{aligned}
\end{equation}
it is a non-convex function, which cannot be directly solved. Its upper bound can be obtained according to \textbf{Lemma 1}.

\noindent\textbf{Lemma 1:} The convex upper bound of $P\left( {{\bf{v}}\left( {i'} \right)} \right)$ is
\begin{equation}
    \begin{aligned}
        {P^{{\textrm{ub}}}}\left( {{\bf{v}}\left( {i'} \right)} \right) &= \frac{{{W^2}}}{{\sqrt 2 \rho S}}{\left( {f_{i'}^n\left( {{{\bf{v}}_h}\left( {i'} \right)} \right)} \right)^{ - \frac{1}{2}}} + W{v_z}\left( {i'} \right)\\
        &+ \frac{{\zeta \rho S}}{8}{\left\| {{{\bf{v}}_h}\left( i' \right)} \right\|^3},
    \end{aligned}
\end{equation}
where 
\begin{equation}
    f_{i'}^n\left( \mathbf{v}_h\left( i' \right) \right)  = f_{i'}^{-n}\left( \mathbf{v}_h\left( i' \right) \right) + 2\left\langle \mathbf{v}_h\left( i' \right), \mathbf{v}_h^n\left( i' \right) \right\rangle - \left\| \mathbf{v}_h^n\left( i' \right) \right\|^2,
\end{equation}
\begin{equation}
    \begin{aligned}
        f_{i'}^{ - n} = {\left( {{{\left\| {{\bf{v}}_h^n\left( {i'} \right)} \right\|}^4} + 4V_h^4} \right)^{ - \frac{1}{2}}}\left( {\left(4V_h^4 + 2\left\langle {{{\bf{v}}_h}\left( {i'} \right),{\bf{v}}_h^n\left( {i'} \right)} \right\rangle \right)}\right.\\ \left.{- {{\left\| {{\bf{v}}_h^n\left( {i'} \right)} \right\|}^2}} \right),
    \end{aligned}
\end{equation}
where ${\bf{v}}_h^n\left( {i'} \right)$ denotes the horizontal velocity of the UAV in the $n$-th iteration.
\newcounter{my3}
\begin{figure*}[!t]
	\normalsize
	\setcounter{my3} 
        {\value{equation}}
	\setcounter{equation}{40}
	\begin{equation}
            \begin{aligned}
                {C_{e,k}}\left( {i'} \right) &= {\log _2}\left( {1 + \frac{{{\rm{Tr}}\left( {{{\bf{W}}_k}\left( {i'} \right)} \right){g_0}{{\left\| {{\bf{q}}\left( {i'} \right) - {{\bf{q}}_e}} \right\|}^{ - {\alpha _2}}}N}}{{{\rm{Tr}}\left( {{\bf{M}}\left( {i'} \right)} \right){g_0}{{\left\| {{\bf{q}}\left( {i'} \right) - {{\bf{q}}_e}} \right\|}^{ - {\alpha _2}}}N + \sigma _e^2\left( {i'} \right)}}} \right)\\
                &  = \underbrace {{{\log }_2}\left( {{\rm{Tr}}\left( {{{\bf{W}}_k}\left( {i'} \right)} \right){g_0}{{\left\| {{\bf{q}}\left( {i'} \right) - {{\bf{q}}_e}} \right\|}^{ - {\alpha _2}}}N + {\rm{Tr}}\left( {{\bf{M}}\left( {i'} \right)} \right){g_0}{{\left\| {{\bf{q}}\left( {i'} \right) - {{\bf{q}}_e}} \right\|}^{ - {\alpha _2}}}N + \sigma _e^2\left( {i'} \right)} \right)}_{C_{e,k}^{1}}\\
                & - \underbrace {{{\log }_2}\left( {{\rm{Tr}}\left( {{\bf{M}}\left( {i'} \right)} \right){g_0}{{\left\| {{\bf{q}}\left( {i'} \right) - {{\bf{q}}_e}} \right\|}^{ - {\alpha _2}}}N + \sigma _e^2\left( {i'} \right)} \right)}_{C_{e,k}^{2}},
            \end{aligned}
        \end{equation}
        \begin{equation}
            {\nabla _{{\bf{q}}\left( {i'} \right)}}C_{e,k}^{{1}}\left( {{\bf{q}}\left( {i'} \right)} \right) = \frac{{ - {\alpha _2}{g_0}N{{\left\| {{\bf{q}}\left( {i'} \right) - {{\bf{q}}_e}} \right\|}^{ - {\alpha _2} - 2}}{{\left( {{\bf{q}}\left( {i'} \right) - {{\bf{q}}_e}} \right)}^T}\left( {{\rm{Tr}}\left( {{{\bf{W}}_k}\left( {i'} \right)} \right) + {\rm{Tr}}\left( {{\bf{M}}\left( {i'} \right)} \right)} \right)}}{{\ln 2\left( {{\rm{Tr}}\left( {{{\bf{W}}_k}\left( {i'} \right)} \right){g_0}{{\left\| {{\bf{q}}\left( {i'} \right) - {{\bf{q}}_e}} \right\|}^{ - {\alpha _2}}}N + {\rm{Tr}}\left( {{\bf{M}}\left( {i'} \right)} \right){g_0}{{\left\| {{\bf{q}}\left( {i'} \right) - {{\bf{q}}_e}} \right\|}^{ - {\alpha _2}}}N + \sigma _e^2\left( {i'} \right)} \right)}},
        \end{equation}
	\setcounter{equation}{\value{my3}}
\hrulefill
\vspace*{4pt}
\end{figure*}

\textit{Proof:} Given the rapid velocity of the UAV, it is reasonable to assume $\left\| {{\bf{v}}_h^n\left( {i'} \right)} \right\| \ge 1$, we have
\begin{equation}
    \begin{aligned}
        2\left\langle {{{\bf{v}}_h}\left( {i'} \right),{\bf{v}}_h^n\left( {i'} \right)} \right\rangle  - {\left\| {{\bf{v}}_h^n\left( {i'} \right)} \right\|^2} &\le
        2\left\| {{\bf{v}}_h^n\left( {i'} \right)} \right\|\left\| {{{\bf{v}}_h}\left( {i'} \right)} \right\|\\ - {\left\| {{\bf{v}}_h^n\left( {i'} \right)} \right\|^2} &\le {\left\| {{\bf{v}}_h^n\left( {i'} \right)} \right\|^2}{\left\| {{{\bf{v}}_h}\left( {i'} \right)} \right\|^2},
    \end{aligned}
\end{equation}
then,
\begin{equation}
    \begin{aligned}
        {\left( {{{\left\| {{\bf{v}}_h^n\left( {i'} \right)} \right\|}^4}\! + \!4V_h^4} \right)^{ - \frac{1}{2}}}\!\left( {4V_h^4\! +\! 2\left\langle {{{\bf{v}}_h}\left( {i'} \right),{\bf{v}}_h^n\left( {i'} \right)} \right\rangle \! - \!{{\left\| {{\bf{v}}_h^n\left( {i'} \right)} \right\|}^2}} \right)\\ \le {\left( {{{\left\| {{\bf{v}}_h^n\left( {i'} \right)} \right\|}^4} \!+\! 4V_h^4} \right)^{ - \frac{1}{2}}} \!\left( {4V_h^4 \!+ \!{{\left\| {{\bf{v}}_h^{}\left( {i'} \right)} \right\|}^2}{{\left\| {{\bf{v}}_h^n\left( {i'} \right)} \right\|}^2}} \right),
    \end{aligned}
\end{equation}
thus,
\begin{equation}
    \begin{aligned}
        f_{i'}^{-n}\left( {{{\bf{v}}_h}\left( {i'} \right)} \right) \le {\left( {{{\left\| {{\bf{v}}_h^n\left( {i'} \right)} \right\|}^4} + 4V_h^4} \right)^{ - \frac{1}{2}}}\\\left( {4V_h^4 + {{\left\| {{\bf{v}}_h\left( {i'} \right)} \right\|}^2}{{\left\| {{\bf{v}}_h^n\left( {i'} \right)} \right\|}^2}} \right).
    \end{aligned}
\end{equation}

To prove $P^{\textrm{ub}}\left( {{\bf{v}}\left( {i'} \right)} \right)$ is the convex upper bound of $P\left( {{\bf{v}}\left( {i'} \right)} \right)$, it is necessary to demonstrate
\begin{equation}
    \begin{aligned}
        {\left( {{{\left\| {{\bf{v}}_h^n\left( {i'} \right)} \right\|}^4} + 4V_h^4} \right)^{ - \frac{1}{2}}}\left( {4V_h^4 + {{\left\| {{\bf{v}}_h^{}\left( {i'} \right)} \right\|}^2}{{\left\| {{\bf{v}}_h^n\left( {i'} \right)} \right\|}^2}} \right) \\\le {\left( {{{\left\| {{\bf{v}}_h^{}\left( {i'} \right)} \right\|}^4} + 4V_h^4} \right)^{\frac{1}{2}}},
    \end{aligned}
\end{equation}
\begin{equation}
    \begin{aligned}
        \left( {4V_h^4 + {{\left\| {{\bf{v}}_h^{}\left( {i'} \right)} \right\|}^2}{{\left\| {{\bf{v}}_h^n\left( {i'} \right)} \right\|}^2}} \right) \le {\left( {{{\left\| {{\bf{v}}_h^{}\left( {i'} \right)} \right\|}^4} + 4V_h^4} \right)^{\frac{1}{2}}}\\{\left( {{{\left\| {{\bf{v}}_h^n\left( {i'} \right)} \right\|}^4} + 4V_h^4} \right)^{\frac{1}{2}}},
    \end{aligned}
\end{equation}
\begin{equation}
    \begin{aligned}
        {\left( {4V_h^4 + {{\left\| {{\bf{v}}_h^{}\left( {i'} \right)} \right\|}^2}{{\left\| {{\bf{v}}_h^n\left( {i'} \right)} \right\|}^2}} \right)^2} \le \left( {{{\left\| {{\bf{v}}_h^{}\left( {i'} \right)} \right\|}^4} + 4V_h^4} \right)\\\left( {{{\left\| {{\bf{v}}_h^n\left( {i'} \right)} \right\|}^4} + 4V_h^4} \right).
    \end{aligned}
\end{equation}
Let ${a^2} = {\left\| {{\bf{v}}_h^{}\left( {i'} \right)} \right\|^4}, {b^2} = 4V_h^4, {c^2} = {\left\| {{\bf{v}}_h^n\left( {i'} \right)} \right\|^4}$, thus,
\begin{equation}
    \left( {{a^2} + {b^2}} \right)\left( {{c^2} + {b^2}} \right) \ge {\left( {ac + {b^2}} \right)^2},
\end{equation}
\begin{equation}
    {a^2}{b^2} + {b^2}{c^2} \ge 2a{b^2}c,
\end{equation}
\begin{equation}
    {b^2}\left( {{a^2} + {c^2} - 2ac} \right) \ge 0.
\end{equation}
Since ${b^2}{\left( {a - c} \right)^2} \ge 0$, the proof of \textbf{Lemma 1} is complete.
$\hfill\blacksquare$

According to Eq. (8), (9) and (16), the communication rate of legitimate users ${R_k}\left( {i'} \right)$ can be transformed as the Eq. (36). It's clear that both functions $R_k^{1}$ and $R_k^{2}$ are concave functions. To transform constraint (22g) into a convex constraint, the following transformation Eq. (37)-(39) is applied to $R_k^{2}$.
\setcounter{equation}{37}
\begin{equation}
\begin{aligned}
    {l_{R_k^{2 }}}\left( {{\bf{q}}\left( {i'} \right)} \right) = R_k^{2 }\left( {{{\bf{q}}^n}\left( {i'} \right)} \right) + {\nabla _{{{\bf{q}}^n}\left( {i'} \right)}}R_k^{2 }\left( {{{\bf{q}}^n}\left( {i'} \right)} \right)\\\cdot \left( {{\bf{q}}\left( {i'} \right) - {{\bf{q}}^n}\left( {i'} \right)} \right),
\end{aligned}
\end{equation}
\begin{equation}
R_k^{\rm{new}}\left( {i'} \right) = R_k^{1} - {l_{R_k^{\rm{2} }}}\left( {{\bf{q}}\left( {i'} \right)} \right).
\end{equation}
Thus, the constraint (22g) is transformed into convex constraint as follows
\begin{equation}
R_k^{\rm{new}}\left( {i'} \right) \ge {R_{\min }}.
\end{equation}
To transform constraint (22h), the following operations can be applied according to the Eq. (41)-(44).
\setcounter{equation}{42}
\begin{equation}
\begin{aligned}
    {l_{C_{e,k}^{1}}}\left( {{\bf{q}}\left( {i'} \right)} \right) = C_{e,k}^{1}\left( {{{\bf{q}}^n}\left( {i'} \right)} \right) + {\nabla _{{{\bf{q}}^n}\left( {i'} \right)}}C_{e,k}^{1}\left( {{{\bf{q}}^n}\left( {i'} \right)} \right)\\ \cdot\left( {{\bf{q}}\left( {i'} \right) - {{\bf{q}}^n}\left( {i'} \right)} \right),
\end{aligned}
\end{equation}

\begin{equation}
C_{e,k}^{\rm{new}}\left( {i'} \right) = {l_{C_{e,k}^{1}}}\left( {{\bf{q}}\left( {i'} \right)} \right) - C_{e,k}^{2}.
\end{equation}
Then, the constraint (22h) is rewritten as 
\begin{equation}
{C_{e,k}^{\rm{new}}}\left( {i'} \right) \le {R_{\max }}.
\end{equation}
Finally, we perform the following manipulations on the objective function $R_k^s\left( {i'} \right)$ to transform it into a concave function
\begin{equation}
R_k^{s,\rm{new}}\left( {i'} \right) = R_k^{\rm{new}}\left( {i'} \right) - C_{e,k}^{\rm{new}}\left( {i'} \right). 
\end{equation}
Therefore, the objective function of  the problem (P4) is successfully transformed into a convex function ${{\varpi _1}} {\left\| {{\bf{q}}\left( {i'} \right) - {{\bf{q}}_B}} \right\|^2} + {{\varpi _2}} {P^{{\textrm{ub}}}}\left( {{\bf{v}}\left( {i'} \right)} \right) -{{\varpi _3}} R_k^{s,\rm{new}}\left( {i'} \right)$. All constraints within the problem (P4) are comprehensively addressed and now ready for direct resolution.
\newcounter{my4}
\begin{figure*}[!t]
	\normalsize
	\setcounter{my4}{\value{equation}}
	\setcounter{equation}{47}
    \begin{equation}
		\begin{aligned}
		    {R_k}\left( i' \right) &= {\log _2}\left( {1 + \frac{{{\rm{Tr}}\left( {{\bf{H}}_k\left( i' \right){{\bf{W}}_k}\left( i' \right)} \right)}}{{{\rm{Tr}}\left( {\sum\limits_{r \in {\cal K}\backslash \left\{ k \right\}} {{\bf{H}}_k\left( i' \right){{\bf{W}}_r}\left( i' \right)} } \right) + {\rm{Tr}}\left( {{\bf{H}}_k\left( i'\right){\bf{M}}\left( i' \right)} \right) + \sigma _k^2\left( i' \right)}}} \right)\\
            &= {\log _2}\left( {{\rm{Tr}}\left( {\sum\limits_{r \in {\cal K}} {{\bf{H}}_k\left( i' \right){{\bf{W}}_r}\left( i' \right)} } \right) + {\rm{Tr}}\left( {{\bf{H}}_k\left( i' \right){\bf{M}}\left( i' \right)} \right)} + \sigma _k^2\left( i' \right) \right) \\
            &- {\log _2}\left( {{\rm{Tr}}\left( {\sum\limits_{r \in {\cal K}\backslash \left\{ k \right\}} {{\bf{H}}_k\left( i' \right){{\bf{W}}_r}\left( i' \right)} } \right) + {\rm{Tr}}\left( {{\bf{H}}_k\left( i' \right){\bf{M}}\left( i' \right)} \right) + \sigma _k^2\left( i' \right)} \right),
		\end{aligned}
	\end{equation}
    \begin{equation}
    \begin{aligned}
        {C_{e,k}}\left( i' \right) &= {\log _2}\left( {1 + \frac{{{\rm{Tr}}\left( {{\bf{G}}_e\left( i' \right){{\bf{W}}_k}\left( i' \right)} \right)}}{{{\rm{Tr}}\left( {{\bf{G}}_e\left( i' \right){\bf{M}}\left( i' \right)} \right) + \sigma _e^2\left( i' \right)}}} \right)\\
        &= {\log _2}\left( {{\rm{Tr}}\left( {{\bf{G}}_e\left( i' \right){\bf{M}}\left( i' \right)} \right) + {\rm{Tr}}\left( {{\bf{G}}_e\left( i' \right){{\bf{W}}_k}\left( i' \right)} \right) + \sigma _e^2\left( i '\right)} \right)\\
        &- {\log _2}\left( {{\rm{Tr}}\left( {{\bf{G}}_e\left( i' \right){\bf{M}}\left( i' \right)} \right) + \sigma _e^2\left( i' \right)} \right).
    \end{aligned}
    \end{equation}
    \begin{equation}
        {\nabla _{{{\bf{W}}_k}\left( {i'} \right)}}{C_{e,k}}\left( {{{\bf{W}}_k}\left( {i'} \right)} \right) = \frac{{{{\bf{G}}_e}\left( {i'} \right)}}{{\ln 2\left( {{\rm{Tr}}\left( {{{\bf{G}}_e}\left( {i'} \right){\bf{M}}\left( {i'} \right)} \right) + {\rm{Tr}}\left( {{{\bf{G}}_e}\left( {i'} \right){{\bf{W}}_k}\left( {i'} \right)} \right) + \sigma _e^2\left( {i'} \right)} \right)}},
    \end{equation}
    \begin{equation}
    \begin{aligned}
       C_{e,k}^{{{\bf{W}}_k}\left( {i'} \right)}\left( {{{\bf{W}}_k}\left( {i'} \right)} \right) &= C_{e,k}^{{{\bf{W}}_k}\left( {i'} \right)}\left( {{\bf{W}}_k^n\left( {i'} \right)} \right) + {\nabla _{{{\bf{W}}_k}\left( {i'} \right)}}{C_{e,k}}\left( {{\bf{W}}_k^n\left( {i'} \right)} \right)\left( {{{\bf{W}}_k}\left( {i'} \right) - {\bf{W}}_k^n\left( {i'} \right)} \right) \\&- {\log _2}\left( {{\rm{Tr}}\left( {{{\bf{G}}_e}\left( {i'} \right){\bf{M}}\left( {i'} \right)} \right) + \sigma _e^2\left( {i'} \right)} \right) ,
    \end{aligned}
    \end{equation}
\setcounter{equation}{\value{my4}}
\hrulefill
\vspace*{4pt}
\end{figure*}

\subsection{Transmit Beamforming Design}
In this subsection, we optimize and design the UAV transmit beamforming given the UAV trajectory and the AN matrix. Thus, this transmit beamforming design problem can be expressed as the problem (P5)
\setcounter{equation}{46}
\begin{subequations}
	\begin{align}
		{\textrm{(P5)}}~~~~&\mathop {\max }\limits_{\mathop {{{\bf{W}}_k}\left( {i'} \right)}} \sum\limits_{i' = i}^{i + {N_p}} {{\varpi _3}\sum\limits_{k \in K} {R_k^{s}\left( {i'} \right)} } ,\nonumber \\
		{\textrm{s.t.}}\qquad &\sum\limits_{k = 1}^K {{\rm{Tr}}\left( {{{\bf{W}}_k}\left( {i'} \right)} \right)}\!  + \!{\rm{Tr}}\left( {{\bf{M}}\left( {i'} \right)} \right) \!+ \!P\left( {{\bf{v}}\left( {i'} \right)} \right) \le {P_{\max }},\\
        &{R_k}\left( {i'} \right) \ge {R_{\min }},\\
		&{C_{e,k}}\left( {i'} \right) \le {R_{\max }},\\
  	&{{\bf{W}}_k}\left( {i'} \right) \succeq 0,\\
		&{\rm{Rank}}\left( {{{\bf{W}}_k}\left( {i'} \right)} \right) = 1.
	\end{align}
\end{subequations}
The inherent non-convexity of the problem, compounded by the rank-1 constraint introduced through matrix dimensionality enhancement, renders the direct solution of this problem exceedingly intricate. To surmount these challenges, we propose transforming the problem into a semi-definite programming (SDP) problem, thereby streamlining the solution process and enhancing its efficiency and feasibility. Specifically, let ${\bf{H}}_k(i') = {\bf{h}}_k^H(i') {\bf{h}}_k(i') \in {\mathbb{C}}^{N \times N}$ and ${\bf{G}}_e(i') = {\bf{g}}_e^H(i') {\bf{g}}_e(i')\in {\mathbb{C}}^{N \times N}$, the transformation of legitimate user communication rates ${R_k}\left( i' \right)$ and eavesdropping rate of potential eavesdropper ${C_{e,k}}\left( i' \right)$ are as Eq. (48) and Eq. (49). Both functions ${R_k}\left( i' \right)$ and ${C_{e,k}}\left( i' \right)$ are concave functions of variable ${{\bf{W}}_k}\left( {i'} \right)$, thus, we proceed to transform the objective function and constraints (47c) and (47e) as delineated below.

First, we apply Eq. (49)-(51) to transform the objective function and constraint (47c) into a concave function and a convex constraint respectively as follows
\setcounter{equation}{51}
\begin{equation}
R_k^{s,{{\bf{W}}_k}\left( {i'} \right)}\left( {i'} \right) = {R_k}\left( {i'} \right) - C_{e,k}^{{{\bf{W}}_k}\left( {i'} \right)}\left( {{{\bf{W}}_k}\left( {i'} \right)} \right),
\end{equation}
\begin{equation}
C_{e,k}^{{{\bf{W}}_k}\left( {i'} \right)}\left( {{{\bf{W}}_k}\left( {i'} \right)} \right) \le {R_{\max }}.
\end{equation}
In addition, for the positive semi-definite matrix ${{\bf{W}}_k}\left( {i'} \right) \in {\mathbb{C}}^{N \times N}$ , ${\rm{Rank}}\left( {{{\bf{W}}_k}\left( {i'} \right)} \right) = 1$, the rank-1 constraint (47e) can be equivalent to the difference between two convex functions, which can be given by
\begin{equation}
{\rm{Rank}}\left( {{{\bf{W}}_k}\left( i \right)} \right) = 1 \Leftrightarrow {\rm{Tr}}\left( {{{\bf{W}}_k}\left( i \right)} \right) - {\left\| {{{\bf{W}}_k}\left( i \right)} \right\|_2} = 0,
\end{equation}
where ${\left\| {{{\bf{W}}_k}\left( {i'} \right)} \right\|_2} = {\sigma _1}\left( {{{\bf{W}}_k}\left( {i'} \right)} \right)$ is spectral norm, and ${\sigma _1}\left( {{{\bf{W}}_k}\left( {i'} \right)} \right)$ represents the first largest singular value of matrix ${{{\bf{W}}_k}\left( i \right)}$. Then, the problem (P5) can be rewritten as the problem (P5.1) as follows
\begin{equation}
	\begin{aligned}
		{\textrm{(P5.1)}}~~~~&\mathop {\max }\limits_{{{\bf{W}}_k}\left( {i'} \right)} \sum\limits_{i' = i}^{i + {N_p}} {\varpi _3}\sum\limits_{k \in K} R_k^{s,{{\bf{W}}_k}\left( {i'} \right)}\left( {i'} \right) \\&~~~~~~~~~- \delta \left( {{\rm{Tr}}\left( {{{{\bf{\bar W}}}_k}\left( {i'} \right)} \right) - {{\left\| {{{{\bf{\bar W}}}_k}\left( {i'} \right)} \right\|}_2}} \right),\\
		& ~~{\textrm{s.t.}}\qquad \textrm{(47a), (47b), (47d), (53)},
	\end{aligned}
\end{equation}
where $\delta$ is penalty factor associated with the rank-1 constraint. The problem (P5.1) is still a non-convex problem because ${\left\| {{\bf{\bar W}}_k}\left( {i'} \right) \right\|_2}$ is convex. Thus, we apply difference-of-convex programming to handle it. Specifically, we need to solve the following problem (P5.2) in the $p$-th iteration,
\begin{equation}
	\begin{aligned}
		{\textrm{(P5.2)}}~~~~&\mathop {\max }\limits_{{{\bf{W}}_k}\left( {i'} \right)} \sum\limits_{i' = i}^{i + {N_p}} {\varpi _3}\sum\limits_{k \in K} R_k^{s,{{\bf{W}}_k}\left( {i'} \right)}\left( {i'} \right) \\&- \delta \left( {\rm{ Tr}}\left( {{\bf{\bar W}}_k}\left( {i'} \right) \right) - \partial \left\langle {\left\| {{\bf{\bar W}}_k}\left( {i'} \right) \right\|_2^{p - 1},{{\bf{\bar W}}_k}\left( {i'} \right) } \right\rangle \right), \\
		& ~~{\textrm{s.t.}}\qquad \textrm{(47a), (47b), (47d), (53)},
	\end{aligned}
\end{equation}
where ${\left\| {{\bf{\bar W}}_k}\left( {i'} \right) \right\|_2^{p - 1}}$ is the solution obtained at the $p - 1$ iteration, and $\partial \left\langle {\left\| {{\bf{\bar W}}_k}\left( {i'} \right) \right\|_2^{p - 1}} \right\rangle $ denotes the subgradient of the spectral norm at the $p - 1$ iteration. The problem (P5.2) is a SDP problem, which can be solved by applying the CVX toolbox \cite{2008CVX}. By solving problem (P5.2) iteratively, we can obtain a rank-1 transmit beamforming matrix.

\newcounter{my5}
\begin{figure*}[!t]
	\normalsize
	\setcounter{my5}{\value{equation}}
	\setcounter{equation}{57}
    \begin{equation}
		\begin{aligned}
		  {R_k}\left( {i'} \right) &= \underbrace {{{\log }_2}\left( {{\rm{Tr}}\left( {\sum\limits_{r \in \kappa } {{{\bf{H}}_k}\left( {i'} \right){{\bf{W}}_r}\left( {i'} \right)} } \right) + {\rm{Tr}}\left( {{{\bf{H}}_k}\left( {i'} \right){\bf{M}}\left( {i'} \right)} \right) + \sigma _k^2\left( {i'} \right)} \right)}_{R_k^{3}}\\
            & - \underbrace {{{\log }_2}\left( {{\rm{Tr}}\left( {{\sum _{r \in \kappa \backslash \{ k\} }}{{\bf{H}}_k}\left( {i'} \right){{\bf{W}}_r}\left( {i'} \right)} \right) + {\rm{Tr}}\left( {{{\bf{H}}_k}\left( {i'} \right){\bf{M}}\left( {i'} \right)} \right) + \sigma _k^2\left( {i'} \right)} \right)}_{R_k^{4}},
		\end{aligned}
	\end{equation}
        \begin{equation}
            {\nabla _{{\bf{M}}\left( {i'} \right)}}R_k^{4}\left( {{\bf{M}}\left( {i'} \right)} \right) = \frac{{{{\bf{H}}_k}\left( {i'} \right)}}{{\ln 2\left( {{\rm{Tr}}\left( {\sum\limits_{r \in \kappa \backslash \{ k\} } {{{\bf{H}}_k}\left( {i'} \right){{\bf{W}}_r}\left( {i'} \right)} } \right) + {\rm{Tr}}\left( {{{\bf{H}}_k}\left( {i'} \right){\bf{M}}\left( {i'} \right)} \right) + \sigma _k^2\left( {i'} \right)} \right)}},
        \end{equation}
        \setcounter{equation}{62}
        \begin{equation}
            {C_{e,k}}\left( {i'} \right) = \underbrace {{{\log }_2}\left( {{\rm{Tr}}\left( {{{\bf{G}}_e}\left( {i'} \right){{\bf{W}}_k}\left( {i'} \right)} \right) + {\rm{Tr}}\left( {{{\bf{G}}_e}\left( {i'} \right){\bf{M}}\left( {i'} \right)} \right) + \sigma _e^2\left( {i'} \right)} \right)}_{C_{e,k}^{3}} - \underbrace {{{\log }_2}\left( {{\rm{Tr}}\left( {{{\bf{G}}_e}\left( {i'} \right){\bf{M}}\left( {i'} \right)} \right) + \sigma _e^2\left( {i'} \right)} \right)}_{C_{e,k}^{4}},
        \end{equation}
        \begin{equation}
            {\nabla _{{\bf{M}}\left( {i'} \right)}}C_{e,k}^{3}\left( {{\bf{M}}\left( {i'} \right)} \right) = \frac{{{{\bf{G}}_e}\left( {i'} \right)}}{{\ln 2\left( {{\rm{Tr}}\left( {{{\bf{G}}_e}\left( {i'} \right){{\bf{W}}_k}\left( {i'} \right)} \right) + {\rm{Tr}}\left( {{{\bf{G}}_e}\left( {i'} \right){\bf{M}}\left( {i'} \right)} \right) + \sigma _e^2\left( {i'} \right)} \right)}},
        \end{equation}
\setcounter{equation}{\value{my5}}
\hrulefill
\vspace*{4pt}
\end{figure*}
{\subsection{AN Matrix Design}
After obtaining the UAV trajectory and transmit beamforming, we solve the AN matrix of the UAV based on the solutions of the above two subproblems in this subsection. The optimization problem can be written as problem (P6), which is expressed as
\setcounter{equation}{56}
\begin{subequations}
	\begin{align}
		{\textrm{(P6)}}~~~~&\mathop {\max }\limits_{\mathop {{{\bf{M}}}\left( {i'} \right)} } \sum\limits_{i' = i}^{i + {N_p}} {{\varpi _3}\sum\limits_{k \in K} {R_k^{s}\left( {i'} \right)} } ,\nonumber \\
		{\textrm{s.t.}}\qquad
		&\sum\limits_{k = 1}^K {{\rm{Tr}}\left( {{{\bf{W}}_k}\left( {i'} \right)} \right)}\!  + \!{\rm{Tr}}\left( {{\bf{M}}\left( {i'} \right)} \right) \!+ \!P\left( {{\bf{v}}\left( {i'} \right)} \right) \le {P_{\max }},\\
        &{R_k}\left( {i'} \right) \ge {R_{\min }},\\
		&{C_{e,k}}\left( {i'} \right) \le {R_{\max }},\\
        &{\bf{M}}\left( {i'} \right) \succeq 0,\\
        &{\rm{Rank}}\left( {{\bf{M}}\left( {i'} \right)} \right) = 1.
	\end{align}
\end{subequations}}Similar to the problem (P5), we implement the the following transformations on the objective function ${R_k^{s}\left( {i'} \right)}$ and constraints (57b), (57c) and (57e).

First, we process the second part of function Eq. (58) (denoted as ${R_k^{4}}$) according to Eq. (59) and (60) as follows.
\setcounter{equation}{59}
\begin{equation}
\begin{aligned}
    R_k^{4,l}\left( {{\bf{M}}\left( {i'} \right)} \right) = R_k^{4}\left( {{{\bf{M}}^n}\left( {i'} \right)} \right) + {\nabla _{{{\bf{M}}^n}\left( {i'} \right)}}R_k^{4}\left( {{{\bf{M}}^n}\left( {i'} \right)} \right)\\\cdot\left( {{\bf{M}}\left( {i'} \right) - {{\bf{M}}^n}\left( {i'} \right)} \right),
\end{aligned}
\end{equation}
\begin{equation}
    R_k^{{\bf{M}}\left( {i'} \right)}\left( {i'} \right) = R_k^{3}\left( {{\bf{M}}\left( {i'} \right)} \right) - R_k^{4,l}\left( {{\bf{M}}\left( {i'} \right)} \right).
\end{equation}
Thus, the constraint (57b) is transformed into a convex constraint as follows
\begin{equation}
R_k^{{\bf{M}}\left( {i'} \right)}\left( {i'} \right) \ge {R_{\min }}.
\end{equation}
Similarly, we handle the constraint (57c) according to Eq. (63)-(66) as follows
\setcounter{equation}{64}
\begin{equation}
\begin{aligned}
    C_{e,k}^{3,l}\left( {{\bf{M}}\left( {i'} \right)} \right) = C_{e,k}^{3}\left( {{{\bf{M}}^n}\left( {i'} \right)} \right) + {\nabla _{{\bf{M}}\left( {i'} \right)}}C_{e,k}^{3}\left( {{{\bf{M}}^n}\left( {i'} \right)} \right)\\\cdot\left( {{\bf{M}}\left( {i'} \right) - {{\bf{M}}^n}\left( {i'} \right)} \right),
\end{aligned}
\end{equation}
\begin{equation}
C_{e,k}^{{\bf{M}}\left( {i'} \right)}\left( {i'} \right) = C_{e,k}^{3,l}\left( {{\bf{M}}\left( {i'} \right)} \right) - C_{e,k}^{4}\left( {{\bf{M}}\left( {i'} \right)} \right).
\end{equation}
Then, the constraint (57c) and objective function are respectively transformed into convex constraint (67) and concave function (68).
\begin{equation}
C_{e,k}^{{\bf{M}}\left( {i'} \right)}\left( {i'} \right) \le {R_{\max }},
\end{equation}
\begin{equation}
R_k^s\left( {i'} \right) = R_k^{{\bf{M}}\left( {i'} \right)}\left( {i'} \right) - C_{e,k}^{{\bf{M}}\left( {i'} \right)}\left( {i'} \right).
\end{equation}
Finally, for the constraint (57e), we adopt the analogous approach used for the constraint (47e), thereby transforming the problem (P6) into the problem (P6.1) as follows
\begin{equation}
	\begin{aligned}
		{\textrm{(P6.1)}}~~~~&\mathop {\max }\limits_{\mathop {{{\bf{M}}}\left( {i'} \right)} } \sum\limits_{i' = i}^{i + {N_p}} {{\varpi _3}\sum\limits_{k \in K} {R_k^{s}\left( {i'} \right)} } \\
        &-\delta\left({\rm{Tr}}\left( {{\bar{\bf{M}}}\left( {i'} \right)}  \right) - \partial \left\langle {\left\|  {{\bar{\bf{M}}}\left( {i'} \right)}  \right\|_2^{p - 1}, {{\bar{\bf{M}}}\left( {i'} \right)} } \right\rangle\right) ,\\
		&~~{\textrm{s.t.}}\qquad  \textrm{(57a), (57d), (62), (67).}
	\end{aligned}
\end{equation}
We can obtain a rank-1 solution by solving the SDP problem (P6.1) iteratively.

\subsection{The Overall MPC Optimization Algorithm}
To enhance the robustness of the system, this paper employs the MPC algorithm for problem-solving. Additionally, to accelerate the convergence of the algorithm, some modifications are introduced to the MPC. The algorithm can be summarized as \textbf{Algorithm 1}.
\begin{algorithm}[htb] 
\caption{The Overall MPC Optimization Algorithm} 
\label{alg:Framwork} 
\begin{algorithmic}[1] 
\REQUIRE ~~\\ 
Starting point ${\bf{q}}_A$, ending point ${\bf{q}}_B$, transmit beamforming
${\bf{w}}_k(0)$, artificial noise ${\bf{m}}(0)$, prediction step size $N_p$,\\ acceptable positional deviation eps, initial number of iterations $\mu  = 0$, acceptable number of iterations $\tau $.\\
\ENSURE ~~\\ 
UAV trajectory, transmit beamforming vector, AN vector\\
\WHILE {$\textrm{norm}({\bf{q}}_A - {\bf{q}}_B) < \textrm{eps}$}
\WHILE {$\mu < \tau$}
\STATE Solve the problem (P4), and update parameters accordingly, i.e., $({\bf{q}}(i'), {\bf{v}}(i'), {\bf{w}}_k(i'), {\bf{m}}(i')) \to ({\bf{q}}(i'+1), {\bf{v}}(i'+1), {\bf{w}}_k(i'), {\bf{m}}(i'))$\\ 
\STATE Solve the problem (P5.2), and update parameters accordingly, i.e., $({\bf{q}}(i' + 1), {\bf{v}}(i' + 1), {\bf{w}}_k(i'), {\bf{m}}(i')) \to ({\bf{q}}(i' + 1), {\bf{v}}(i' + 1), {\bf{w}}_k(i' + 1), {\bf{m}}(i'))$\\ 
\STATE Solve the problem (P6.1), and update parameters accordingly, i.e., $({\bf{q}}(i' + 1), {\bf{v}}(i' + 1), {\bf{w}}_k(i' + 1), {\bf{m}}(i')) \to ({\bf{q}}(i' + 1), {\bf{v}}(i' + 1), {\bf{w}}_k(i' + 1), {\bf{m}}(i' + 1))$\\
\STATE $\mu \rightarrow {\mu + 1}$\\
\ENDWHILE
\STATE Temporal rolling and upgrate the initial parameters for the subsequent iteration\\
\STATE $\mu \rightarrow{0}$, ${\bf{q}}_A \to {\bf{q}}_A + {\bf{v}}(0) \times t_c + {\textrm{disturb}}$, $i \rightarrow{i + 1}$\\
\ENDWHILE
\RETURN UAV trajectory, transmit beamforming vector, AN vector\\ 
\end{algorithmic}
\end{algorithm}

\subsection{Convergence Analysis}
The convergence of the proposed \textbf{Algorithm 1} can be proved as follows.

We define ${\bf{v}}^p, {\bf{W}}^p, {\bf{M}}^p$ as the $p$-th iteration solution of the problem (P4), (P5.2) and (P6.1). Herein, the objective function is denoted by 
$\mathcal{E}({\bf{q}}^p,\mathbf{v}^p, \mathbf{W}^p, \mathbf{M}^p)$. In the solution of problem (P4), since the transmit beamforming vector and AN vector can be obtained for given ${\bf{W}}^p, {\bf{M}}^p$. Hence, we have 
\begin{equation}
    {\mathcal{E}(\mathbf{q}^{p+1}, \mathbf{v}^{p+1}, \mathbf{W}^p, \mathbf{M}^p)} \le {\mathcal{E}(\mathbf{q}^p, \mathbf{v}^p, \mathbf{W}^p, \mathbf{M}^p)}.
\end{equation}
In the solution of the problem (P5.2), since the UAV trajectory design and AN vector can be obtained for given ${\bf{q}}^{p+1}, {\bf{v}}^{p+1},{\bf{M}}^p$. Hence, we also have 
\begin{equation}
    {\mathcal{E}(\mathbf{q}^{p+1}, \!\mathbf{v}^{p+1},\! \mathbf{W}^{p+1}, \!\mathbf{M}^p)}\! \le\! {\mathcal{E}(\mathbf{q}^{p+1}, \!\mathbf{v}^{p+1}, \!\mathbf{W}^p,\! \mathbf{M}^p)}.
\end{equation}
Finally, in the solution of the problem (P6.1), since the UAV trajectory design and transmit beamforming vector can be obtained for given ${\bf{q}}^{p+1}, 
 {\bf{v}}^{p+1}, {\bf{W}}^{p+1}$. Thus, we have 
\begin{equation}
    {\mathcal{E}(\mathbf{q}^{p+1}, \mathbf{v}^{p+1}, \mathbf{W}^{p+1},\mathbf{M}^{p+1})} \le {\mathcal{E}(\mathbf{q}^{p+1}, \mathbf{v}^{p+1}, \mathbf{W}^{p+1}, \mathbf{M}^p)}.
\end{equation}
Based on the above, we can obtain
\begin{equation}
    {\mathcal{E}(\mathbf{q}^{p+1}, \mathbf{v}^{p+1}, \mathbf{W}^{p+1}, \mathbf{M}^{p+1})} \le {\mathcal{E}(\mathbf{q}^{p}, \mathbf{v}^p, \mathbf{W}^p, \mathbf{M}^p)}.
\end{equation}
It shows that the value of the objective function after each iteration of\textbf{ Algorithm 1} is non-increasing. Meanwhile, the objective function value of the problem (P3) has a lower bound, so the convergence of \textbf{Algorithm 1} can be guaranteed.

\section{Numerical Results}
In this section, we utilize the CVX toolbox in MATLAB to conduct simulations of the model, with the specific simulation parameters presented in the Table I \cite{34}.
\newcommand{\tabincell}[2]{\begin{tabular}{@{}#1@{}}#2\end{tabular}}
\begin{table}
	\begin{center}
		\renewcommand{\arraystretch}{}
		\caption{Simulation Parameters}
		\label{T1}
		{\begin{tabular}{|c|c|}
			\hline
			\textbf{Parameters}&\textbf{Value}\\
			\hline
			{Time slot interval $t_c$}&10s\\
			\hline
			{$h_k$ path loss exponent $\alpha_1$}&2.3\\
			\hline
			{$g_k$ path loss exponent $\alpha_2$}&2.5\\
			\hline
			{Weight of UAV $W$}&39.2kg\\
			\hline
			{Air density $\rho$}&1.225$\mathrm{kg/m^3}$\\
			\hline
			{Total area of the UAV rotor disc $S$}&1$\mathrm{m^2}$\\
			\hline
			{Profile drag coefficient $\zeta$}&0.08\\
			\hline
			{Maximum horizontal velocity $V_{max}$}&120m/s\\
			\hline
			{Maximum vertical velocity $U_{max}$}&30m/s\\
			\hline
			{Maximum acceleration $a_{max}$}&8$\rm{m/s^2}$\\
			\hline
			{Minimum flight altitude $Z_{min}$}&500m\\
			\hline
			{Maximum flight altitude $Z_{max}$}&900m\\
			\hline
		\end{tabular}}
	\end{center}
\end{table}

\begin{figure}[t]
	\centering  
	\includegraphics[scale=0.42]{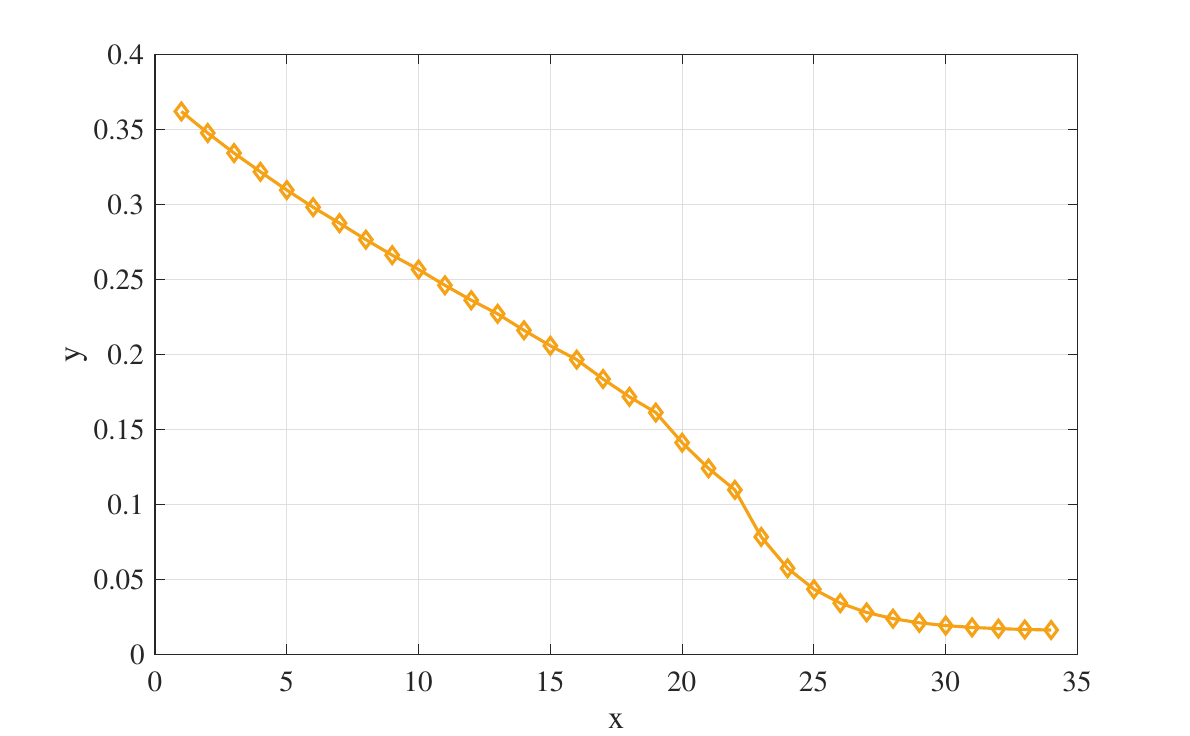}
	\caption{Convergence behavior of the proposed algorithm.}  %
	\label{Figure 1} 
\end{figure}
Fig. 2 illustrates the convergence of the algorithm. The curve indicates that as iterations progress, the objective function steadily decreases, with the gradient of the curve approaching zero. This trend signifies that the objective function converges to a specific value, thereby demonstrating the algorithm's convergence.

\begin{figure}[t]
	\centering  
	\includegraphics[scale=0.42]{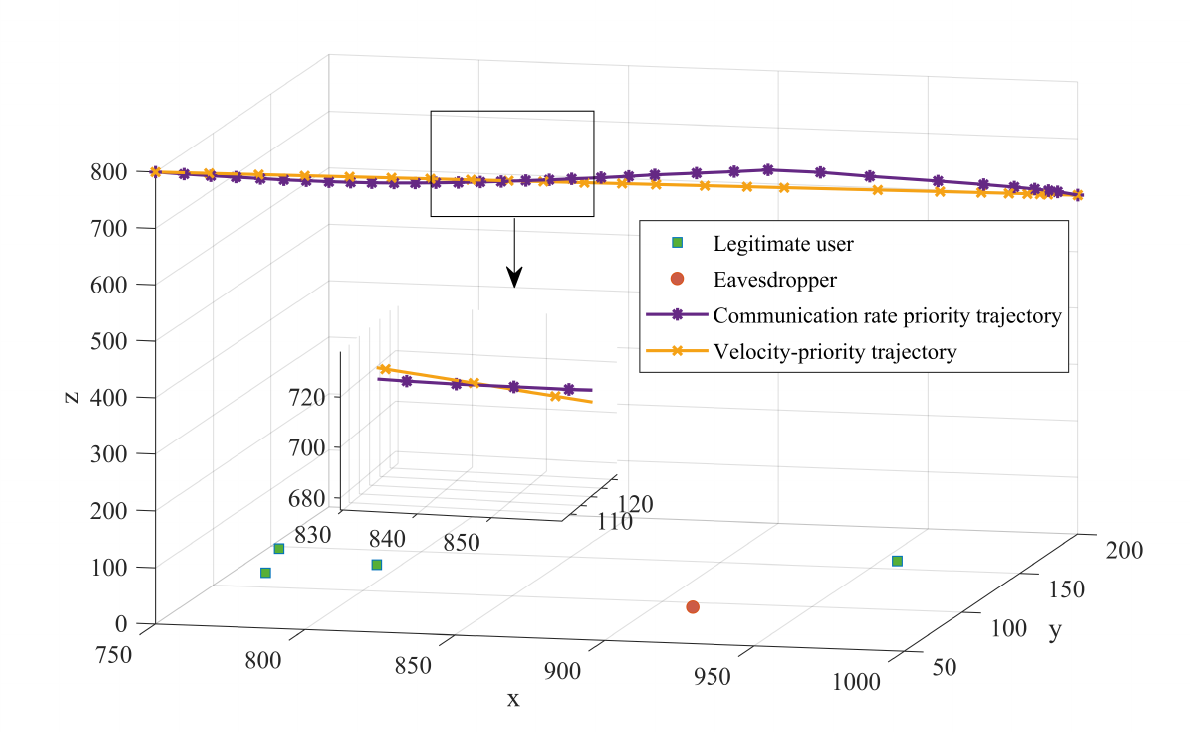}
	\caption{Different metrics-prioritized trajectory for UAV.}  %
	\label{Figure 1} 
\end{figure}
Fig. 3 shows the trajectories of the UAV under two distinct scenarios: speed prioritization and secure communication rate prioritization, with annotations indicating the positions of legitimate users and potential eavesdropper. In the absence of disturbances, the speed-prioritized trajectory is represented as a straight line from the starting point to the endpoint, as evidenced by the trajectory depicted in the figure. Conversely, when prioritizing secure communication rates, the UAV descends upon detecting a legitimate user to enhance communication services. In proximity to potential eavesdropper, the UAV ascends to a higher altitude to diminish the eavesdropper's communication rate, thereby ensuring the quality of user service, as illustrated by the trajectory corresponding to secure communication rate prioritization.

\begin{figure}[t]
	\centering  
	\includegraphics[scale=0.42]{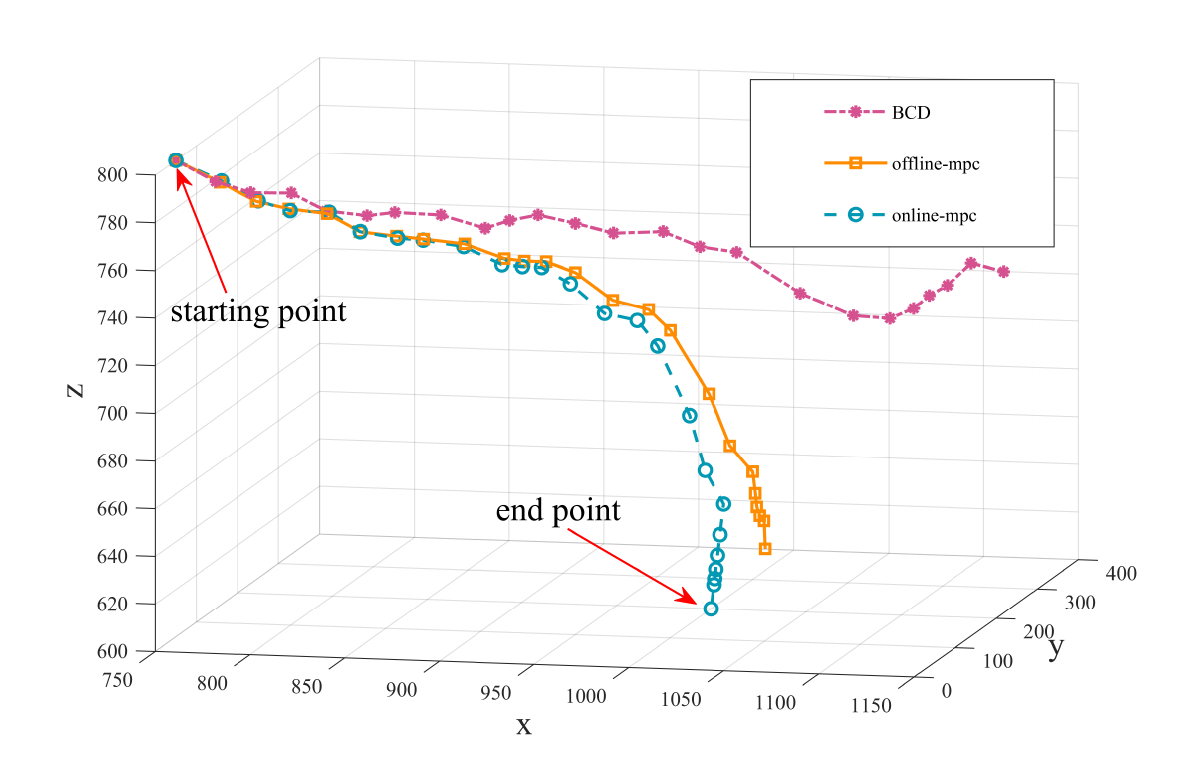}
	\caption{UAV trajectories under different benchmarks.}  
	\label{Figure 1} 
\end{figure}
Fig. 4 presents the trajectories of the UAV under the influence of disturbances, as governed by the BCD algorithm, the offline MPC algorithm, and the online MPC algorithm. It is evident that the application of the BCD algorithm can result in significant deviations of the UAV trajectory from the predetermined endpoint due to disturbances. In contrast, the MPC algorithms exhibit notable robustness, enabling real-time corrections of these deviations. When disturbances are entirely known, both the offline and online MPC algorithms demonstrate equivalent performance. However, discrepancies between the actual and known disturbances may lead to a decline in the performance of the offline MPC algorithm.

\begin{figure}[t]
	\centering  
	\includegraphics[scale=0.42]{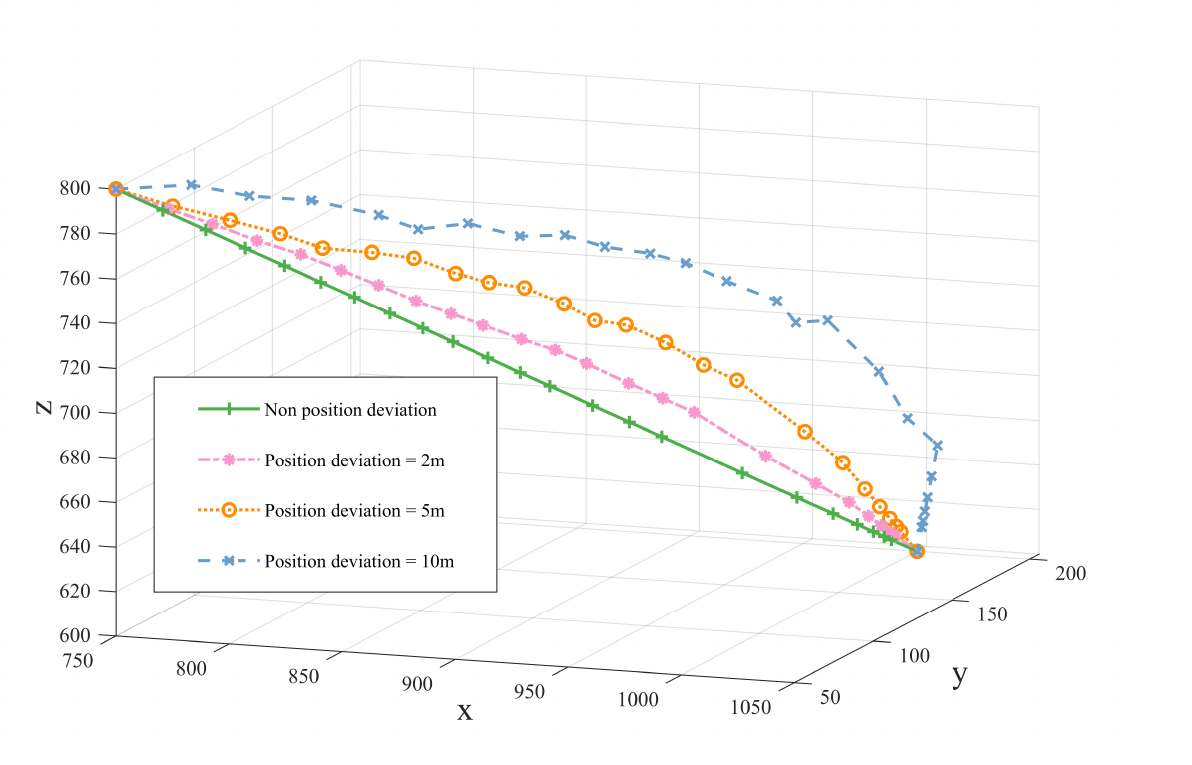}
	\caption{UAV trajectories under varying disturbance conditions.}  %
	\label{Figure 1} 
\end{figure}
Fig. 5 illustrates the trajectories of the UAV under varying intensities of disturbance. The figure demonstrates that, under the influence of these disturbances, the UAV's trajectories cluster around the trajectory observed in the absence of disturbances. This indicates that the UAV is capable of effectively correcting deviations caused by disturbances and successfully reaching the predetermined endpoint, thereby validating the robustness of the MPC algorithm and enhancing the stability of the system.

\begin{figure}[t]
	\centering  
	\includegraphics[scale=0.42]{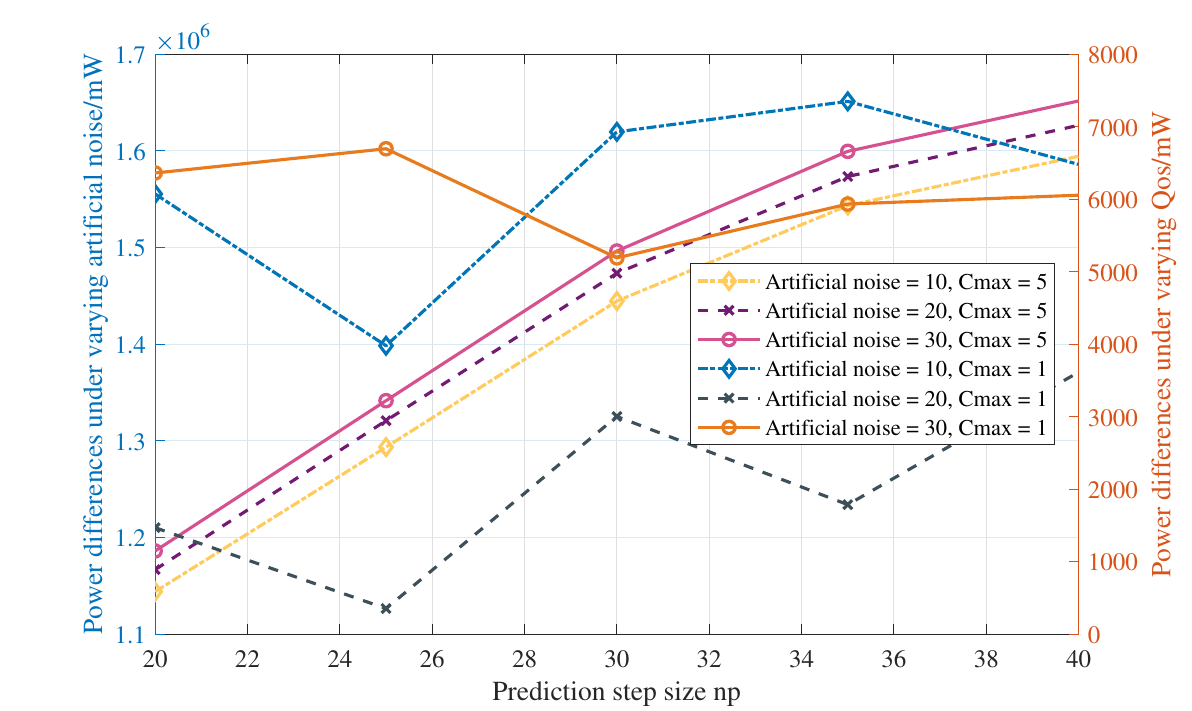}
	\caption{Power consumption for varying prediction step size.}  %
	\label{Figure 1} 
\end{figure}
Fig. 6 depicts the total power required for the UAV to emit varying intensities of AN vector at different prediction horizons, alongside the impact of changes in the upper threshold of potential eavesdropper's communication rates on power consumption. As the prediction horizon increases, the transmit beamforming vector allocation by the UAV rises, resulting in an escalation of the allocated power. However, the trajectory remains relatively stable, leading to negligible variations in flight power loss and a corresponding increase in total power with the prediction horizon. Furthermore, an increase in AN vector intensity results in greater power consumption due to enhanced transmit beamforming. When the upper threshold for potential eavesdropper's communication rates is lowered to improve user service quality, the transmit beamforming intensity allocated to potential eavesdropper is consequently diminished, leading to a reduction in power consumption compared to the transmission of AN vector at equivalent intensities.

\begin{figure}[t]
	\centering  
	\includegraphics[scale=0.42]{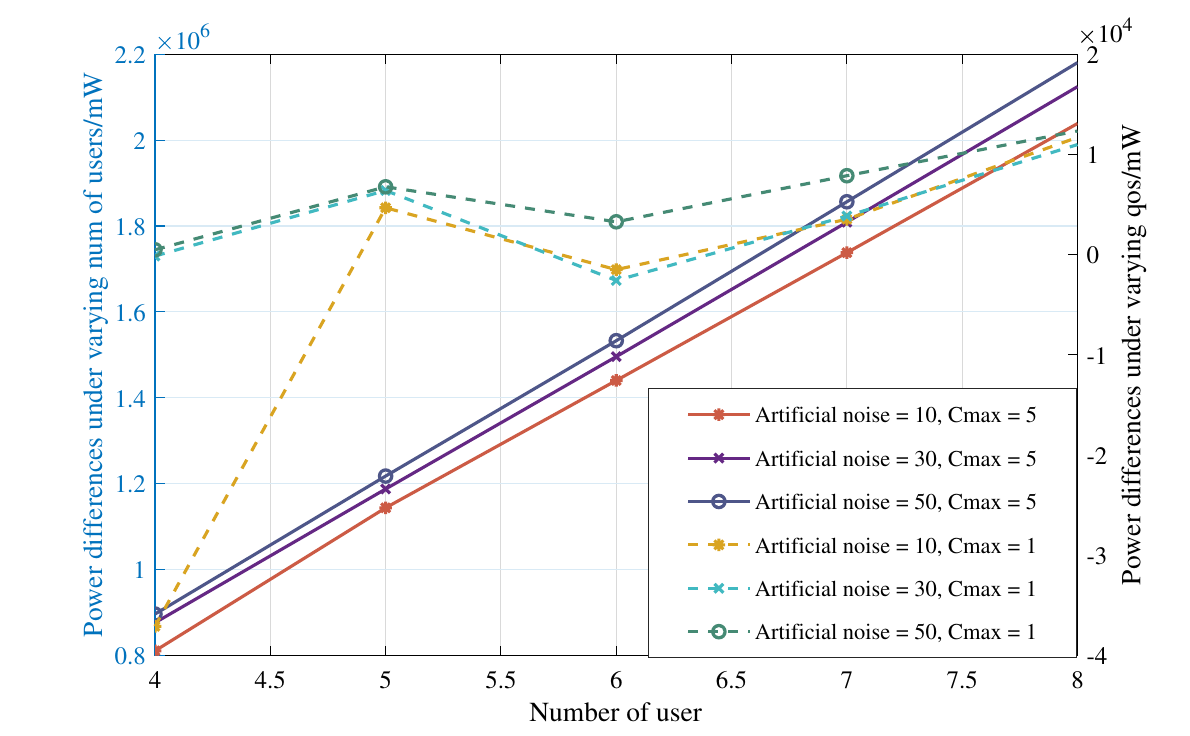}
	\caption{Power consumption for varying numbers of users.}  %
	\label{Figure 1} 
\end{figure}
Fig. 7 illustrates the total power required for the UAV to emit varying intensities of AN vector as a function of the number of users, alongside the power variations resulting from adjustments to the upper threshold of potential eavesdropper's communication rates. As the number of users increases, the UAV allocates transmit beamforming vector to a greater number of targets, consequently resulting in an increase in the allocated power. Additionally, with the enhancement of AN vector intensity, the associated power consumption rises in tandem due to intensified transmit beamforming. When the upper threshold for potential eavesdropper's communication rates is lowered to enhance user service quality, the transmit beamforming allocated to potential eavesdropper is inherently diminished, leading to a reduction in power consumption compared to the transmission of AN vector at equivalent intensities.

\section{Conclusions}
This paper has investigated the UAV-enabled secure communication system, and the maneuverability of UAV can provide guarantee for the safety performance of the system. Specifically, a control problem of minimizing UAV flight path and power consumption while maximizing secure communication rate over infinite horizon has been formulated by jointly optimizing UAV trajectory, transmit beamforming vector, and AN vector. Initially, the problem has been transformed and decomposed into three subproblems. In the first subproblem, the trajectory optimization of the UAV has been achieved by solving a transformed convex optimization problem. Subsequently, in the second and third subproblem, the transmit beamforming and AN vector have been optimized by solving the SDP problems. Furthermore, this paper has introduced the difference-of-convex programming to handle non-convex rank-1 constraints. Ultimately, the three subproblems have been solved iteratively until the UAV reaches the predetermined destination range. Additionally, an analysis of the convergence of the proposed algorithm has been conducted. Simulation results have indicated that the algorithm enhances system robustness, enabling UAVs to provide communication services to users while ensuring communication security in the presence of disturbances, with potential applications in complex environment rescue scenarios.

\bibliographystyle{IEEEtran}
\bibliography{reference}

\vfill

\end{document}